\documentclass[aps,prd,twocolumn,showpacs,preprintnumbers,nofootinbib,float,superscriptaddress]{revtex4-1}

\usepackage{epsfig}
 
\usepackage[dvipsnames]{xcolor}
\usepackage{float,amsmath,amssymb}
\usepackage{graphicx}
\usepackage{url}
\usepackage{bbold}
\usepackage[utf8]{inputenc}
\usepackage{physics}
\newcommand{\lqcd}{\Lambda_{\mathrm{QCD}}}
\newcommand{\as}{{\alpha_{\mathrm{s}}}}

\newcommand{\xt}{{\mathbf{x}_\perp}}

\newcommand{\yt}{{\mathbf{y}_\perp}}
\newcommand{\bt}{{\mathbf{b}_\perp}}
\newcommand{\zt}{{\mathbf{z}_\perp}}
\newcommand{\bti}{{\mathbf{b}_{\perp,i}}}

\newcommand{\nc}{{N_\mathrm{c}}}
\newcommand{\nf}{N_\mathrm{f}}
\newcommand{\jpsi}{$\mathrm{J}/\psi$ }

\usepackage[breaklinks,colorlinks,citecolor=citcolor,urlcolor=blue,linkcolor=lcolor]{hyperref}
\definecolor{lcolor}{rgb}{0.5,0,0}
\definecolor{citcolor}{rgb}{0,0.3,0.0}

\begin{document}

\title{Impact of QCD Energy Evolution on Observables in Heavy-Ion Collisions}

\author{Heikki M\"antysaari}
\affiliation{Department of Physics, University of Jyv\"askyl\"a, P.O. Box 35, 40014 University of Jyv\"askyl\"a, Finland}
\affiliation{Helsinki Institute of Physics, P.O. Box 64, 00014 University of Helsinki, Finland}

\author{Bj\"orn Schenke}    
\affiliation{Physics Department, Brookhaven National Laboratory, Upton, NY 11973, USA}

\author{Chun Shen}
\affiliation{Department of Physics and Astronomy, Wayne State University, Detroit, Michigan 48201, USA}

\author{Wenbin Zhao}
\affiliation{Institute of Particle Physics and Key Laboratory of Quark and Lepton Physics (MOE), Central China Normal University, Wuhan, 430079, Hubei, China}

\begin{abstract}
    We study how the inclusion of energy dependence as dictated by quantum chromodynamic (QCD) small-$x$ evolution equations affects key observables in ultra-relativistic heavy-ion collisions. Specifically, we incorporate JIMWLK evolution into the IP-Glasma framework, which serves as the initial condition for a simulation pipeline that includes viscous relativistic hydrodynamics and a hadronic afterburner. This approach enables a consistent modeling of highly energetic nuclei across varying Bjorken-$x$ values, which are relevant for different collision energies and rapidity regions. In comparison to the standard IP-Glasma setup without small-$x$ evolution, we observe pronounced changes in particle multiplicities and spectral distributions, especially in smaller systems and at the highest available energies. We further explore effects on anisotropic flow observables and correlations between mean transverse momentum and elliptic flow. Our findings underscore the critical role of nonlinear QCD evolution in accurately modeling the early stages of heavy-ion collisions, as well as its implications for extracting transport properties of the quark-gluon plasma. 
\end{abstract}

\maketitle

\section{Introduction}
Ultra-relativistic heavy-ion collisions are a powerful tool to probe nuclear matter under extreme conditions of temperature and density. 
Most collision energies at the Relativistic Heavy Ion Collider (RHIC) are high enough to facilitate the transition to the quark-gluon plasma (QGP)  \cite{Muller:2012zq, Harris:2024aov}. Much effort has been devoted to exploring the properties of the QGP, which can lead to a deeper understanding of many-body quantum chromodynamics (QCD). One important insight we have gained is that the matter produced in collisions at RHIC and the Large Hadron Collider (LHC) behaves like a nearly perfect fluid~\cite{Heinz:2013th,Gale:2013da,Jeon:2015dfa,Romatschke:2017ejr,Heinz:2024jwu}, meaning that its viscosity over entropy density ratio is among the lowest realized in nature.

Center-of-mass energies accessible at RHIC and LHC span several orders of magnitude, demanding that the hydrodynamic modeling incorporate a controlled collision energy dependence. This energy dependence is typically embedded within the modeling of the system's initial conditions. In the case of the Monte Carlo Glauber framework, parameters such as the nucleon-nucleon interaction cross section and an overall normalization factor are tuned to reflect variations in particle production with collision energy \cite{Miller:2007ri}. In the Eskola-Kajantie-Ruuskanen-Tuominen (EKRT) model~\cite{Kuha:2024kmq} the energy dependence results from the behavior of nuclear parton distribution functions that were fitted to data. Similarly, the IP-Glasma approach \cite{Schenke:2012wb}, which relies on the color glass condensate (CGC) formalism~\cite{Iancu:2003xm,Garcia-Montero:2025hys} -- an effective field theory of QCD -- captures energy dependence through adjustments in the saturation scale, guided by the constraints on parameters of the IPSat model~\cite{Kowalski:2003hm} obtained from deep inelastic scattering data~\cite{Rezaeian:2012ji}.

While these approaches have proven effective, successfully reproducing a wide range of observables across different collision energies \cite{Schenke:2020mbo}, the growing complexity of experimental measurements and the increasing precision of data call for a more refined treatment of the $\sqrt{s}$-dependence in modeling the initial conditions for hydrodynamic simulations.

In this work, and in a companion letter~\cite{Mantysaari:2025tcg}, we expand our analysis of the effects of incorporating the JIMWLK equations~\cite{Jalilian-Marian:1997qno,Iancu:2000hn,Mueller:2001uk,Lappi:2012vw,Cali:2021tsh} into the IP-Glasma model. Including the JIMWLK evolution allows us to describe the nuclei at the appropriate $x$-values, representing the longitudinal momentum fraction of the participating partons, and approximately given by $\langle p_T \rangle/\sqrt{s}$, where $\langle p_T \rangle$ is the average transverse momentum of produced particles. 

In \cite{Mantysaari:2025tcg} we studied how predicted charged particle multiplicities and transverse momenta in different collision systems were affected by including JIMWLK evolution as compared to the original IP-Glasma model, where the energy dependence follows by adjusting the saturation scale $Q_s$ to the appropriate $x$-dependent value. Here, besides providing more details on the combined IP-Glasma+JIMWLK model, we extend our analysis to more differential observables, including various anisotropic flow measurements and correlations between mean transverse momentum and elliptic flow. In particular, we quantify how the evolution in the spatial structure of light and heavy nuclei affects the bulk observables in high-energy collisions. 

We note that JIMWLK evolution has been included in heavy-ion initial state descriptions in the past to study the rapidity dependence of the initial geometry~\cite{Schenke:2016ksl,Schenke:2022mjv,McDonald:2023qwc,Mantysaari:2024qmt}. As in \cite{Mantysaari:2025tcg}, we focus on midrapidity results, and leave an extension to simulations in 3+1D, where the JIMWLK evolution will affect both the collision energy and rapidity dependence of the collisions, to future work.

This paper is structured as follows. The IP-Glasma+JIMWLK initial state is described in detail in Sec.~\ref{sec:ebe-geometry}. The space-time evolution, consisting of the early-time Glasma phase coupled to relativistic hydrodynamics in the QGP phase and a hadronic afterburner, is presented in Sec.~\ref{sec:cym_hydro}. Results for heavy (Au+Au and Pb+Pb), intermediate (O+O and Ne+Ne), and small (p+Pb)  collision systems are shown in Sec.~\ref{sec:results}.

\section{IP-Glasma initial state with JIMWLK evolution}
\label{sec:ebe-geometry}

To model the spatial configuration of colliding nuclei, initial collision, and pre-equilibrium evolution, we employ the IP-Glasma framework \cite{Gale:2012rq,Schenke:2012wb}. In its standard formulation---widely used to describe a range of RHIC and LHC data \cite{Schenke:2020mbo}---the IP-Glasma model encodes the impact parameter $\bt$ and Bjorken-$x$ dependence via a local saturation scale $Q_s^2(x, \bt)$. The saturation scale itself is extracted from the IPSat model~\cite{Kowalski:2003hm}, calibrated against HERA deep inelastic scattering data~\cite{Rezaeian:2012ji,H1:2009pze}.

In contrast to this traditional approach, our current work introduces a dynamically evolving energy dependence by numerically solving the JIMWLK evolution equations. This shift brings a key distinction: while the IP-Sat-based saturation scale yields a static geometric profile with respect to $x$, the JIMWLK evolution captures the growth and smoothing of hadronic and nuclear profiles as $x$ decreases. This results in increasingly diffuse and spatially extended color charge distributions at small-$x$, a behavior observed in prior studies~\cite{Mantysaari:2022sux,Schlichting:2014ipa}, and consistent with the $x$-dependence of generalized parton distribution functions that were constrained with experimental data~\cite{Moutarde:2018kwr}. Furthermore, our approach does not rely on any extracted parametrization of $Q_s^2$, as we instead solve the perturbative evolution equations for the high-energy nuclear structure.

Our initialization of the small-$x$ evolution uses the same nuclear structure inputs as in earlier versions of the IP-Glasma model~\cite{Gale:2012rq,Schenke:2012wb,Schenke:2020mbo,Mantysaari:2022sux}. At the starting scale $x = x_0 = 0.01$, the color field configuration of each nucleon is encoded in Wilson lines $V(\xt)$, defined on a discrete transverse grid. These Wilson lines are obtained from the Yang-Mills equations, using sampled color charges as external currents moving with the speed of light. The color charges are obtained from the McLerran-Venugopalan (MV) model~\cite{McLerran:1993ni}, where they are described as random local Gaussian variables with a correlator
\begin{align}
    g^2 \langle \rho^a&(x^-,\xt) \rho^b(y^-,\yt)\rangle \notag\\ = &\delta^{ab} \delta^{(2)}(\xt-\yt) \delta(x^- - y^-)
    \times g^4 \lambda_A(x^-).
    \label{eq:MV}
\end{align}
Boldface coordinates denote 2D vectors in the plane transverse to the beam direction. The variables  $x^-$ and $y^-$ are lightcone coordinates, defined as $x^{\pm} = (t\pm z)/\sqrt{2}$, where $t$ is the time and $z$ the longitudinal (beam) direction.
The color charge density is $\mu^2 = \int \dd{x^-} \lambda_A(x^-)$, which (at the initial condition at $x=0.01$) is related to the saturation scale $Q_s^2$ extracted from the IPsat model: $g^2 \mu(x=x_0, \bt) = c Q_s(x=x_0,\bt)$. The proportionality factor $c = g^2\mu/Q_s$ is treated as a free parameter in the model. The Wilson lines are given by
\begin{equation}
    V(\xt) = P_- \exp \left\{ -ig \int \dd{x^-} \frac{\rho^a(x^-,\xt) t^a}{\nabla^2_{\xt} - \tilde{m}^2} \right\},
\end{equation}
Here $\tilde{m}$, which is also a free parameter, is an infrared regulator suppressing long distance Coulomb tails.

The saturation scale $Q_s^2$ in the IPsat model depends on the transverse spatial profile of the target, represented by the thickness function $T(\bt)$. For nucleons, this profile is modeled as a superposition of $N_q$ Gaussian-shaped hot spots~\cite{Mantysaari:2016ykx, Mantysaari:2020axf, Mantysaari:2025ltq}, where each hot spot’s position $\bti$ is sampled from a Gaussian distribution with width $B_{qc}$. The center of mass of the $N_q=3$ hot spots is shifted to the origin after sampling. The resulting expression for the nucleon thickness function is 
\begin{align}
    T(\bt)&= \sum_{i=1}^{N_q} T_q(\bt-\bti), \quad \text{where} \\
    T_q(\bt) &= \frac{1}{2\pi B_{q}} \exp\left(-\frac{\bt^2}{2B_q}\right).
\end{align}
Here, $B_q$ defines the transverse width of each hotspot and serves as a tunable parameter in the model.

For nuclear targets, nucleon positions are distributed according to the nuclear density profile and the total thickness function is obtained by summing over the contributions from all constituent nucleons. For large nuclei we employ Woods-Saxon profiles (with a minimum distance of 0.9 fm between the nucleons following~\cite{Broniowski:2010jd, Schenke:2020mbo}), and for smaller nuclei we use results from more sophisticated  nuclear structure calculations. 
The  configurations of nucleons for $^{16}$O and $^{20}$Ne are obtained from   \textit{ab initio} Projected Generator Coordinate Method (PGCM) calculations~\cite{Giacalone:2024ixe,Giacalone:2024luz}, and enforcing two protons and two neutrons to
sit close to $\alpha$-cluster centers.~\footnote{We note that in Ref.~\cite{Mantysaari:2025tcg} we used the Variational Monte-Carlo (VMC) \cite{Carlson:1997qn} for the oxygen configurations.} The deuteron wave function is sampled from the simple Hulthen wave function with the parameters experimentally determined in Ref.~\cite{Miller:2007ri}.

The model further incorporates event-by-event variations in the local saturation scale, implemented through independent $Q_s$ fluctuations for each hot spot. This is done because having such fluctuations is preferred by low-$|t|$ incoherent diffractive $\mathrm{J}/\psi$ production at HERA~\cite{Mantysaari:2016ykx,Mantysaari:2025idf}. Specifically, the normalization of each $T_q$ is scaled by a stochastic factor $p_i$ drawn from a log-normal distribution
\begin{equation}
\label{eq:qsfluct}
P\left( \ln p_i \right) = \frac{1}{\sqrt{2\pi}\sigma} \exp \left[- \frac{(\ln p_i)^2}{2\sigma^2}\right]\,.
\end{equation}
The fluctuation width $\sigma$ is a free parameter of the model, and the normalization of sampled values is adjusted by dividing by the mean value $e^{\sigma/2}$ to preserve the overall scale.

The dependence of Wilson lines on the energy scale, or equivalently on Bjorken-$x$, can be determined by numerically solving the JIMWLK evolution equation~\cite{Cali:2021tsh,Mueller:2001uk}. Adopting the formulation presented in Refs.~\cite{Lappi:2012vw,Mantysaari:2018zdd}, the evolution of the Wilson line $V(\xt)$ with rapidity $y = \ln(x_0/x)$ can be expressed in a form where updates involve left and right multiplication of $V(\xt)$ as follows:\footnote{We follow the ``square root'' running coupling prescription and omit the inclusion of the coupling constant in the noise term, differing slightly from the implementation in Ref.~\cite{Mantysaari:2018zdd}.}
\begin{multline}
\label{eq:jimwlk}
V_\xt(y+\dd y) = \exp \bigg\{ -i \frac{\sqrt{ \dd y}}{\pi} \int \dd[2]{\zt} \sqrt{\as(|\zt|)}  \\ 
\times  \mathbf{K}_{\xt-\zt} \cdot (V_\zt \boldsymbol{\xi}_\zt V^\dagger_\zt) \bigg\}\\
\times V_\xt(y) \exp \left\{ i\frac{\sqrt{ \dd y}}{\pi} \int \dd[2]\zt \sqrt{\as(|\zt|)}  \mathbf{K}_{\xt-\zt} \cdot \boldsymbol{\xi}_\zt \right\},
\end{multline}
where $\boldsymbol{\xi}_\zt = (\xi_{\zt,1}^a t^a, \xi_{\zt,2}^a t^a)$, with 1,2 labeling the spatial directions, and $\xi_{\zt,i}^a$ is a local random Gaussian noise. 

The JIMWLK kernel, which describes the probability distribution for soft gluon emissions, is given by
\begin{equation}
K_\xt^i =  m|\xt| K_1(m|\xt|) \frac{x^i}{\boldsymbol{x}_\perp^2}.
\end{equation}
Following Refs.~\cite{Schlichting:2014ipa,Mantysaari:2018zdd}, we introduced an infrared regulator $m$ to tame the long-range Coulomb tails that would otherwise cause the proton size to grow faster than allowed by the Froissart bound. This regulator is a free parameter in the model.
In the limit $m\to 0$ the above kernel reduces to the perturbative gluon emission kernel $x^i/\boldsymbol{x}_\perp^2$.

The strong coupling constant is evaluated in coordinate space as
\begin{equation}
\as(r) = \frac{12\pi}{(11\nc - 2\nf) \ln \left[ \left(\frac{\mu_0^2}{\lqcd^2}\right)^{1/\zeta} + \left(\frac{4}{r^2\lqcd^2}\right)^{1/\zeta} \right]^\zeta}\,,
\label{eq:alphas}
\end{equation}
with parameters $\mu_0 = 0.28\,\mathrm{GeV}$, $\zeta = 0.2$, and $\nf = 3$ as in Ref.~\cite{Mantysaari:2022sux}. The scale parameter $\lqcd$, which controls the running of the coupling in coordinate space, is treated as a free parameter in the model.

In this work, we compare two different strategies for modeling the energy dependence of the small-$x$ structure of the proton or heavier nucleus:
\begin{enumerate}
\item \emph{IP-Glasma+JIMWLK}: This approach follows the setup of Ref.~\cite{Mantysaari:2022sux}, where the Wilson lines are obtained by solving the full JIMWLK evolution starting from the parametrized initial condition at $x = 0.01$ to $x=\langle p_T\rangle e^{\pm y}/\sqrt{s_\mathrm{NN}}$, with $\langle p_T\rangle$ being the charged hadron mean transverse momentum. We use $\left\langle p_T\right\rangle=0.5$ GeV at RHIC kinematics and $\left\langle p_T\right\rangle=0.65$ GeV at the LHC. The relative variation in $\left\langle p_T\right\rangle$ from RHIC to LHC is much smaller than the changes in collision energy when determining the $x$ values, such that the precise choice of $\langle p_T \rangle$ values does not play a significant role.
In this work, we consider midrapidity kinematics, $y=0$.
The model parameters are constrained by fitting to coherent and incoherent \jpsi photoproduction data from H1, ZEUS, ALICE, CMS, and LHCb, as described in more detail below.

\item \emph{$Q_s(x)$}: 
As a baseline model, we freeze the model parameters in Setup 1 to their values at the initial $x=0.01$. The energy dependence is included by taking the proton saturation scale to be $x$-dependent as described by the IPsat model. 
This setup is equivalent to the usual IP-Glasma model as used e.g.~in \cite{Schenke:2020mbo}, except that we follow the original implementation of IP-Glasma, with $x$ determined as described above. We do not determine $Q_s$ iteratively by solving $x = x(\xt) = Q_s(x,\xt)/\sqrt{s_\mathrm{NN}}$, which has been done e.g.~in \cite{Schenke:2020mbo}.  This is done to be able to compare more directly with the JIMWLK setup, in which a position dependent $x(\xt)$ would require saving the evolution for all $x$ and evaluate different $x$ slices depending on the position. This would require significantly larger computational and memory resources. 
\end{enumerate}
By construction, these setups are identical at $x=0.01$. 

The non-perturbative model parameters control the proton's color charge distribution ($c, \tilde m$) and its event-by-event geometric fluctuations ($B_{qc},B_q,\sigma$) at the initial $x=x_0$. Furthermore, the coordinate space running coupling scale $\lqcd$ controls the speed of the JIWMLK evolution. 
These parameters have been determined in Ref.~\cite{Mantysaari:2022sux} by fitting to a comprehensive set of data on exclusive $\gamma + p \rightarrow \mathrm{J}/\psi + p$ production. This includes measurements from HERA~\cite{H1:2005dtp,H1:2013okq,ZEUS:2002wfj} and from ultraperipheral $p$ + Pb collisions at the LHC~\cite{Bertulani:2005ru,Klein:2019qfb,ALICE:2014eof,ALICE:2018oyo,LHCb:2014acg,LHCb:2018rcm}, spanning a wide range in Bjorken-$x$ from approximately $10^{-2}$ down to $10^{-6}$. The minimum transverse distance between the two hot spots inside a nucleon, $d_{q, \mathrm{min}}$, is set to $0.254$ fm following Ref.~\cite{Mantysaari:2022ffw}.
With that, all initial state parameters, except the overall normalization factor for the energy density profile, have been constrained by exclusive vector meson production data. 
 The values of $\Lambda_{\rm QCD}$ shown in Table 1 are relatively small, which is a result of it being defined in coordinate space (as opposed to the usual momentum space scale) and the fact that we adjust $\Lambda_{\rm QCD}$ directly to control the evolution speed. In comparison, other works (see, e.g., Refs. \cite{Albacete:2010sy,Lappi:2013zma}) introduce an additional constant $C^2$ (which appears in Eq. (7) of Ref.~\cite{Lappi:2013zma}: $\sim (4/r^2) C^2/\Lambda_{\rm QCD}^2$). There, $C^2$ is of order 10, effectively achieving the same effect as choosing a smaller $\Lambda_{\rm QCD}$. The need for a small $\Lambda_{\rm QCD}$ originates from the leading order evolution being too fast: In \cite{Lappi:2016fmu} it was demonstrated that next-to-leading order effects slow down evolution, something we have to capture effectively using the value of $\Lambda_{\rm QCD}$.

We also note that recently an updated analysis that  also takes into account vector meson photoproduction in $\gamma+\mathrm{Pb}$ collisions, and provides uncertainty estimates for the model parameters, has been presented in Ref.~\cite{Mantysaari:2025ltq}. In the future, it would be beneficial to quantify how uncertainties in the non-perturbative initial state parameters propagate to final state observables in heavy ion collisions.

\begin{table}[tb] 
\caption{Initial state parameter values used in this work. \label{tab:initparams}}
\centering
\begin{tabular}{|c|c|c|c|c|}
\hline
$c$ & $\tilde{m}$ & $B_{qc}$ & $B_{q}$ & $\sigma$  \\
\hline
0.643 & 0.4 GeV & 3.3 GeV$^{-2}$ & 0.3 GeV$^{-2}$ & 0.7 \\
\hline
\hline
 $m$ &  $\Lambda_{\rm QCD}$ & $K_{\rm JIMWLK}$&$K_{\rm Q_s(x), RHIC}$ &$K_{\rm Q_s(x), LHC}$ \\
\hline
 0.4 GeV & 0.04 GeV &0.115 &0.106 & 0.095\\
\hline
\end{tabular}
\end{table}

Open-source implementations of the IP-Glasma model and the JIMWLK evolution are publicly available~\cite{ipglasma_code,jimwlk_code}. In addition, a modified version of IP-Glasma coupled to JIMWLK evolution can be accessed at~\cite{ipglasma_jimwlk_code}.

\section{Space-time evolution}
\label{sec:cym_hydro}
As mentioned above, the gluon fields after the collision are obtained in the standard way~\cite{Krasnitz:1998ns,Krasnitz:1999wc,Krasnitz:2000gz,Schenke:2012wb} from the Wilson lines determined with one of the above prescriptions. Further, the early-time evolution in the Glasma phase up to $\tau_0=0.4$ fm/c is obtained by solving the source-free Yang-Mills equations as implemented in the IP-Glasma model. 
The energy-momentum tensor $T^{\mu\nu}$ of the system is computed from the gluon fields at $\tau=\tau_0$, providing the initial condition for relativistic viscous hydrodynamic simulations, which are performed using MUSIC~\cite{Schenke:2010nt,Schenke:2010rr,Paquet:2015lta}.

Since the strong coupling constant factors out of the classical Yang-Mills evolution equations (see discussion in~\cite{Schenke:2020mbo}), we determine an overall normalization factor $K$ for the energy-momentum tensor $T^{\mu\nu}$ by matching to the measured charged hadron multiplicity in the 0--5\% most central Au+Au collisions at $\sqrt{s_\mathrm{NN}} = 200$~GeV. Once fixed, the JIMWLK evolution predicts the energy dependence of particle production across different collision energies.
This stands in contrast to the conventional Glauber-based approach, where the normalization is typically re-tuned separately for each center-of-mass energy.
We note that running coupling effects can be expected to result in $K$ having a logarithmic dependence on the center-of-mass energy. As running coupling corrections are included in the JIMWLK evolution employed in this work, analogous corrections could, in principle, be applied to the $K$ factor. This would introduce an additional dependence on the choice of running coupling scheme and is therefore left for future work.   Note that $K$ in  Table.~\ref{tab:initparams} should not be of order 1 but of order $1/g^2$, with the coupling $g$ defined in Eq.~(\ref{eq:MV}). Since the classical calculation of the early-time dynamics is independent of $g$, we set $g=1$ in the numerical simulations. Then, $K$ reintroduces the correct dependence of the initial energy density on $\alpha_s$. The preferred value $K\simeq0.1$ corresponds to an effective coupling $g\approx3$. See the discussion in Ref.~\cite{Schenke:2020mbo} for more details on this subject. We further remark that in principle $K$ is expected to depend on the collision energy as it is scale dependent.

Focusing exclusively on midrapidity observables, we assume boost invariance in the hydrodynamic evolution. We employ an equation of state derived from lattice QCD calculations~\cite{Bazavov:2014pvz,Moreland:2015dvc}, and include both shear and bulk viscous effects. These are treated using the Denicol-Niemi-Molnar-Rischke (DNMR) second-order viscous hydrodynamic formalism~\cite{Denicol:2012cn}, which incorporates spatial gradient corrections up to second order. We do not incorporate an intermediate pre-equilibrium stage. This could be done most directly using the existing KoMPoST framework \cite{Kurkela:2018wud}, however there are reasons this is not easily achievable: 1) KoMPoST suffers from numerical instabilities in the presence of large pressure gradients, especially in peripheral and small collision systems. 2) The dynamics of KoMPoST remain conformal, which leads to a longer conformal pre-equilibrium evolution compared to the Glasma phase alone. This results in stronger effects on final-state observables from the discontinuity introduced by switching from the conformal system to one described by the lattice QCD equation of state.

The transition from the fluid to hadronic degrees of freedom is performed on an isothermal hypersurface defined by a constant energy density $e_\mathrm{sw} = 0.18~\mathrm{GeV/fm}^3$, using the Cooper-Frye particlization procedure with viscous corrections implemented via Grad's 14-moment method~\cite{Huovinen:2012is,Shen:2014vra,Zhao:2022ugy}. The resulting hadrons are subsequently evolved in the dilute hadronic phase using the UrQMD transport model~\cite{Bass:1998ca,Bleicher:1999xi}, which accounts for hadronic scatterings and decays.

The hydrodynamic evolution is sensitive to the transport properties of the medium, particularly the temperature-dependent shear ($\eta$) and bulk ($\zeta$) viscosities. In this work, we employ the following parametrizations for $\eta/s$ and $\zeta/s$:
\begin{align}
    \frac{\eta}{s}(T) &= \eta_0 + b(T - T_0) \Theta(T_0 - T) \\
    \frac{\zeta}{s}(T) &= \frac{(\zeta/s )_{\max}\Lambda^2}{\Lambda^2+ \left( T-T_0\right)^2},
    \Lambda = w_{\zeta} \left[1 + \lambda_{\zeta} {\rm sign} \left(T{-}T_0\right) \right].
\end{align}
Here $s$ and $T$ are the entropy density and temperature, respectively.
We tune the parameters to reproduce experimental data on charged hadron multiplicity~\cite{PHENIX:2003iij}, mean transverse momentum~\cite{STAR:2008med}, and azimuthal flow coefficients~\cite{STAR:2015mki} measured in Au+Au collisions at $\sqrt{s_\mathrm{NN}} = 200$~GeV. The IP-Glasma+JIMWLK setup is used for this tune.
The purpose of this work is to quantify the effects of including the small-$x$ JIMWLK evolution in the initial state description. Consequently, the transport properties are constrained using the RHIC data only, as there is only moderate evolution from initial $x_0=0.01$ to RHIC kinematics corresponding to $x=0.0025$.
The obtained parameters are $\eta_0 = 0.14$, $T_0 = 0.18$ GeV, $b = -4$ GeV$^{-1}$, $(\zeta/s )_{\max} = 0.123$, $\lambda_{\zeta}=-0.12$, and $ w_{\zeta}= 0.05$ GeV. 

\section{Results}
\label{sec:results}
In this section, we present results for various collision systems, starting with the largest heavy ion collisions (Au+Au at RHIC and Pb+Pb at LHC), then moving to smaller systems (O+O and Ne+Ne), and finally exploring p+Pb collisions. Besides the energy dependence of charged particle multiplicities, we focus on anisotropic flow observables and how they are affected by the different treatments of the energy dependence.

\subsection{Heavy ion collisions}

\begin{figure}[tb]
    \centering    \includegraphics[width=\columnwidth]{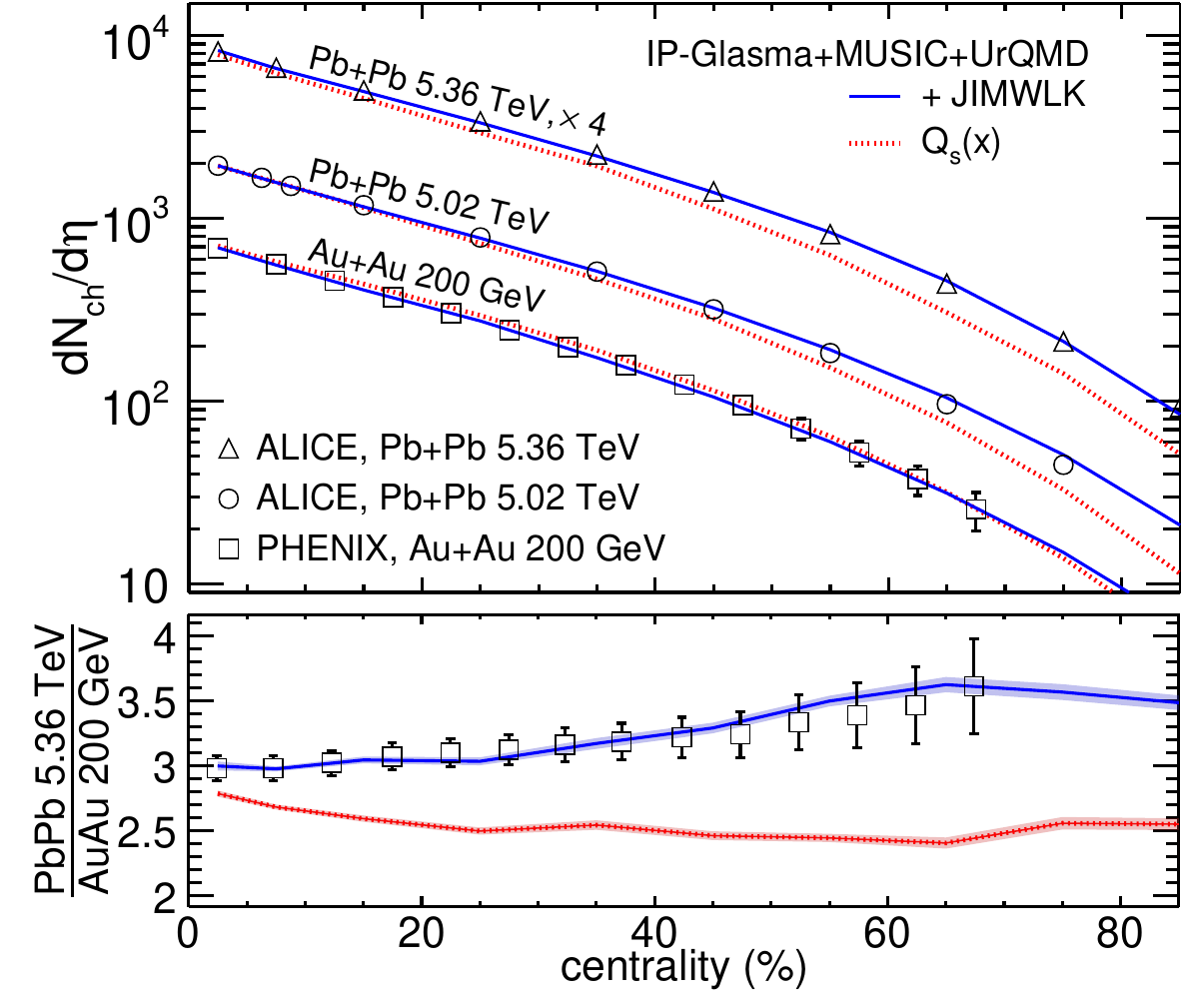}
    \caption{Charged hadron multiplicity distributions in Pb+Pb collisions at the LHC  5.02 TeV and 5.36 TeV, and Au+Au collisions at RHIC 200 GeV energies. The experimental data are taken from \cite{ALICE:2015juo,ALICE:2025cjn,PHENIX:2004vdg}. }
    \label{fig:dNch_cent_PbPb}
\end{figure}

Charged particle multiplicity distributions as a function of centrality in $\mathrm{Pb}+\mathrm{Pb}$ collisions at $\sqrt{s}=5.02$ and $\sqrt{s}=5.36$ TeV are shown along with those for Au+Au collisions at $\sqrt{s}=200$ GeV in Fig.~\ref{fig:dNch_cent_PbPb}. Throughout this work, we show results obtained using both initial state descriptions described in Sec.~\ref{sec:ebe-geometry}: our main setup referred to as ``+JIMWLK'', and the baseline setup ``$Q_s(x)$''. We remind the reader that all parameters of the hydrodynamic simulation were fixed to describe the 200 GeV Au+Au data, with all other systems and energies being true model predictions.

Both prescriptions for the initial state energy dependence reproduce the multiplicity in the most central collisions.
On the other hand, in more peripheral Pb+Pb collisions beyond a centrality of $\sim 20\%$, inclusion of the JIMWLK evolution results in significantly flatter multiplicity distributions. 
We interpret this to be due to the geometry evolution included in the ``+JIMWLK'' setup, which modifies the overlap by blurring the edges of the incoming nuclei,  and consequently changes the shape of the multiplicity distribution with centrality. 
This geometry evolution has been previously studied e.g. in  Refs.~\cite{Schlichting:2014ipa,Mantysaari:2018zdd,Mantysaari:2022sux}. We have checked that the differences in the viscous entropy production during the late-time hydrodynamic phase are not significant.

To further quantify how sensitive the multiplicity distribution is to the initial state energy evolution, we calculate the ratio between the LHC Pb+Pb and RHIC Au+Au multiplicities. Centrality dependence of this ratio is shown in the bottom panel of Fig.~\ref{fig:dNch_cent_PbPb}. With the JIMWLK evolution describing the initial state energy evolution, the increasing trend of this ratio as a function of centrality can be reproduced, while the ratio obtained in the ``$Q_s(x)$'' setup is too flat in centrality.

\begin{figure}[tb]
    \centering
    \includegraphics[width=\columnwidth]{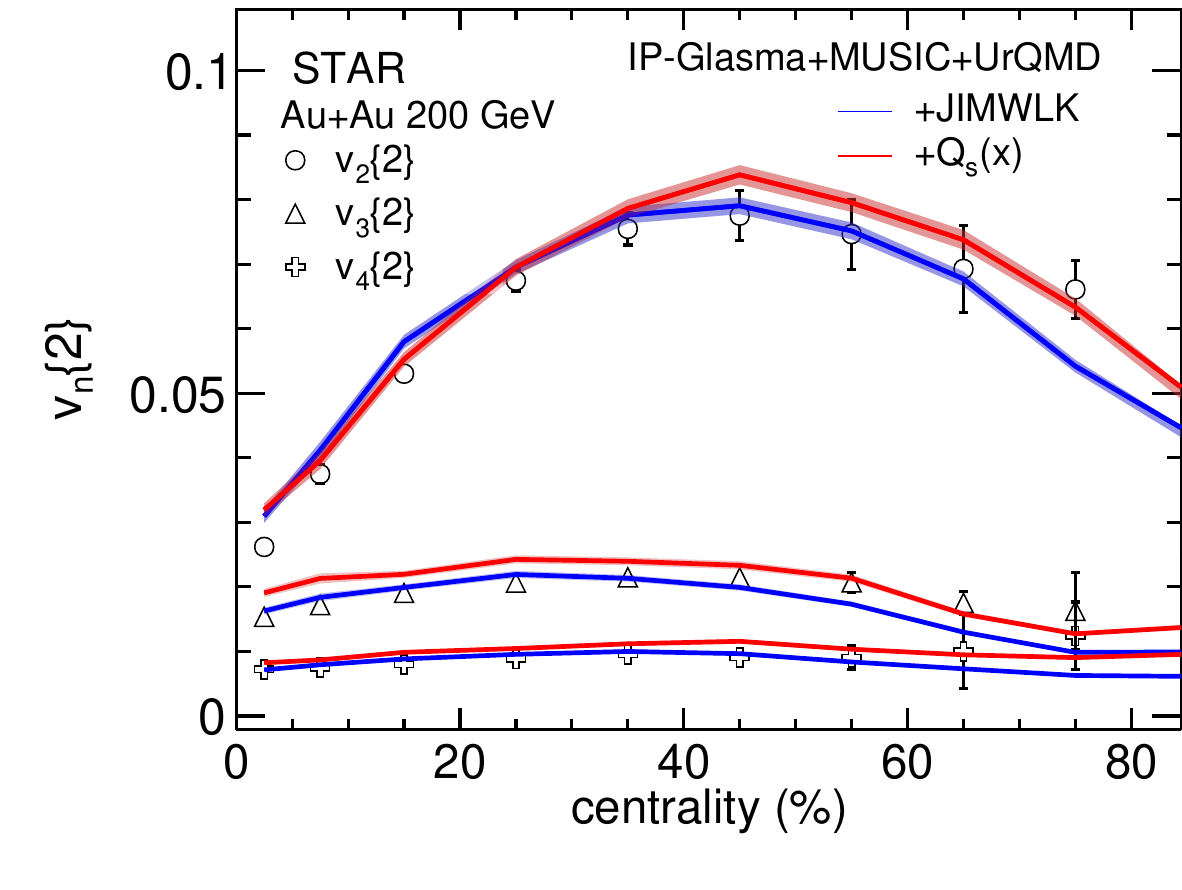}
    \caption{Anisotropic flow $v_n$ in Au+Au collisions 200 GeV. The experimental data are from \cite{STAR:2016vqt,STAR:2017idk,STAR:2019zaf}.}
    \label{fig:AuAu_vn}
\end{figure}

\begin{figure}[tb]
    \centering
    \includegraphics[width=0.9\columnwidth]{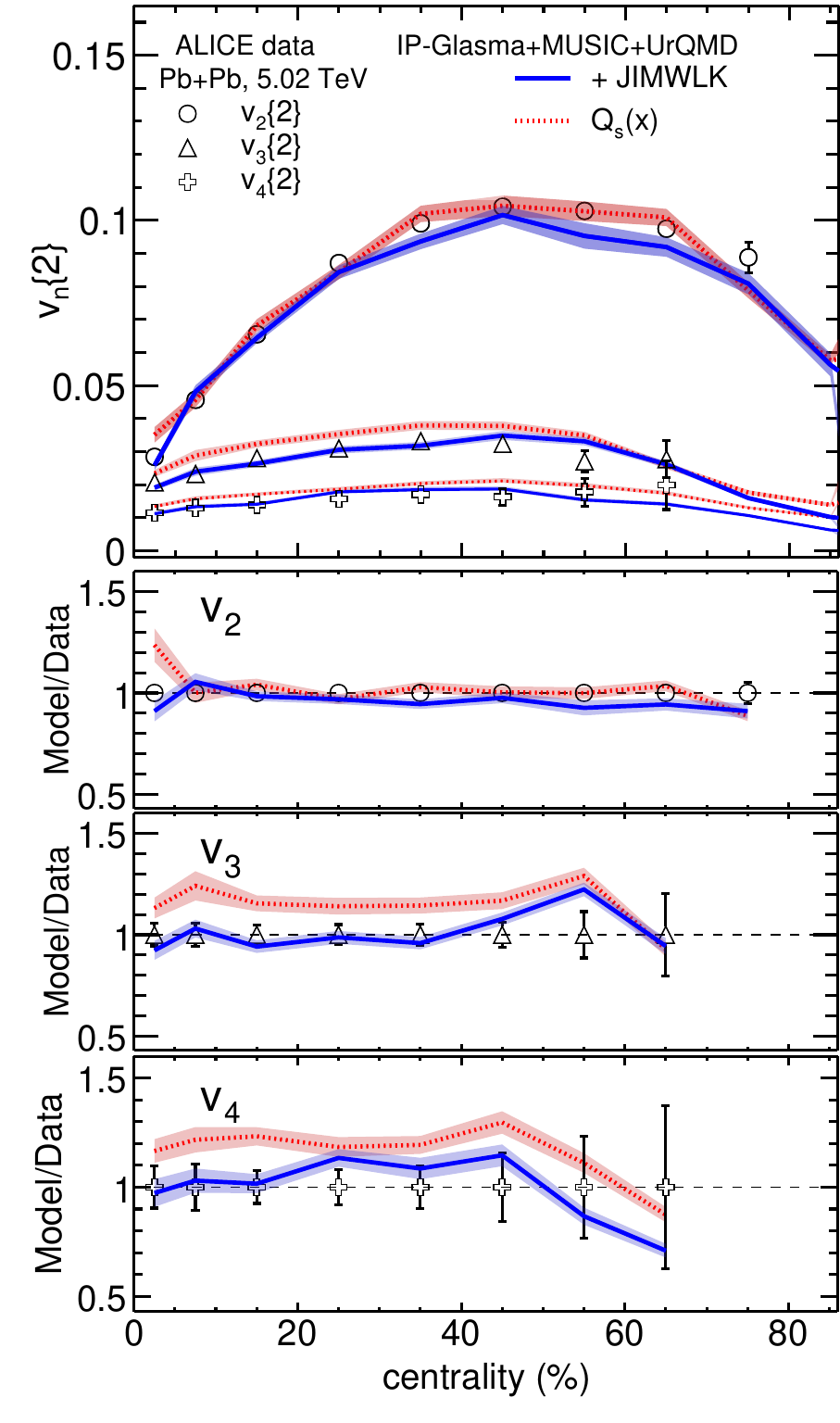}
    \caption{Collective flow $v_n$ in Pb + Pb collisions 5.02 TeV. The experimental data are from \cite{ALICE:2016ccg}. }
    \label{fig:PbPb_vn}
\end{figure}

\begin{figure}[tb]
    \centering
    \includegraphics[width=0.9\columnwidth]{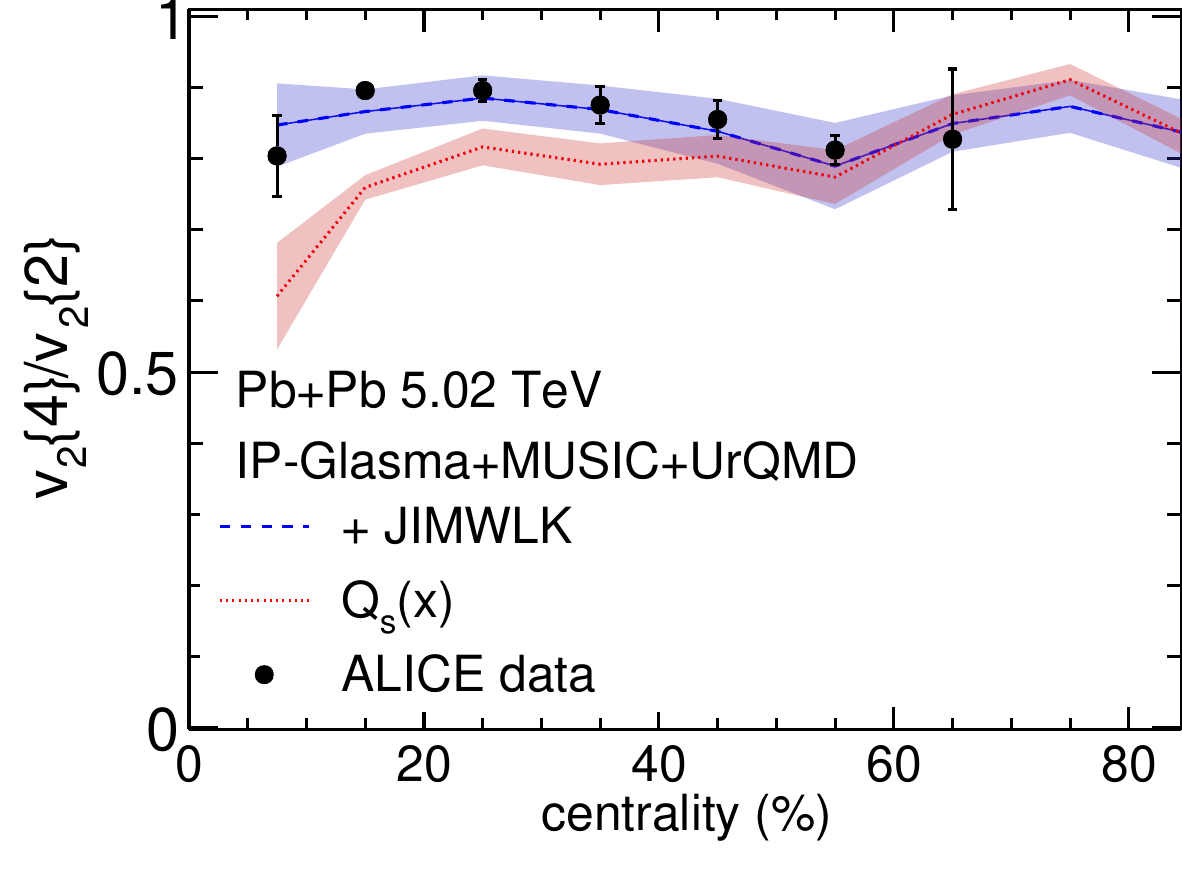}
    \caption{The  $v_2\{4\}/v_2\{2\}$ ratio in Pb + Pb collisions at 5.02 TeV. The experimental data are from \cite{ALICE:2016ccg}. }
    \label{fig:PbPb_v24_v22_ratio}
\end{figure}

The effect of JIMWLK evolution on anisotropic flow coefficients was not studied in the companion letter~\cite{Mantysaari:2025tcg}. To set the stage, we first show our results for $v_2\{2\}, v_3\{2\}$  and $v_4\{2\}$ in Au+Au collisions at 200 GeV center-of-mass energy in Fig.\,\ref{fig:AuAu_vn}.
Note that shear and bulk viscosities have been adjusted to describe this data  in the ``+JIMWLK" setup, and kept the same in the $Q_s(x)$ setup - see discussion in Sec.~\ref{sec:cym_hydro}.
There is only moderate evolution from the initial $x=0.01$ to the typical midrapidity value at RHIC, $x=0.0025$. 
Yet, there is some difference between our results for $v_2\{2\}$ and $v_2\{4\}$ for centralities larger than 40\% as well as for $v_3\{2\}$ for all centralities. In all cases the +JIMWLK result is lower than the corresponding $Q_s(x)$ result. This can again be explained by the smoothing effect of the JIMWLK evolution on the initial state geometry, which reduces the initial eccentricities, as we have checked explicitly.

Similar trends are seen for Pb+Pb collisions at 5.02 TeV: The centrality dependence of the flow harmonics $v_2\{2\}, v_3\{2\}$  and $v_4\{2\}$ is shown in Fig.~\ref{fig:PbPb_vn}. Again, the JIMWLK evolution is found to suppress all flow harmonics compared to the $Q_s(x)$ setup. Interestingly, the differences are similar to those observed in the lower energy Au+Au collisions, suggesting that the dominant effect from the small-$x$ evolution is the smoothening of the initially lumpy nuclear substructure early in the evolution. Faster evolution at larger $x$ is expected as saturation effects that slow down the evolution become more prominent as $x$ decreases.
The ALICE data~\cite{ALICE:2016ccg} is found to prefer the setup with the JIMWLK evolution included. Our main observation is that final state bulk observables are sensitive to the details of the small-$x$ evolution in the initial state.

To further quantify the differences between the two energy evolution schemes, we calculate the $v_2\{4\}/v_2\{2\}$ ratio in LHC kinematics. The obtained ratios as a function of centrality are compared to the ALICE data~\cite{ALICE:2016ccg} in Fig.~\ref{fig:PbPb_v24_v22_ratio}. Here, only the ``+JIMWLK'' setup results in an approximately constant ratio, consistent with the ALICE data. On the other hand, an increasing trend with centrality, especially around small centrality, is obtained in the $Q_s(x)$ setup. This qualitative difference can be explained as follows. 
Because $v_2\{2\}$ increases with flow fluctuations, while $v_2\{4\}$ decreases (which has been shown explicitly e.g.~for Gaussian fluctuations), this ratio is a measure of fluctuations, with smaller values implying larger fluctuations~\cite{Voloshin:2008dg}. As the JIMWLK evolution leads to smoother nuclei, initial state fluctuations are suppressed, leading to a larger ratio than in the $Q_s(x)$ setup.

Results for the average transverse momentum at different energies were already shown in \cite{Mantysaari:2025tcg}. Similarly to the anisotropic flow coefficients, the smoothing (and growth) of the geometry in the +JIMWLK setup led to reduced $\langle p_T\rangle$ in the Pb+Pb system at 5.02 TeV and an improved description of the ALICE data~\cite{ALICE:2018hza}.

\subsection{O+O and Ne+Ne collisions}

\begin{figure}[tb]
    \centering
    \includegraphics[width=\columnwidth]{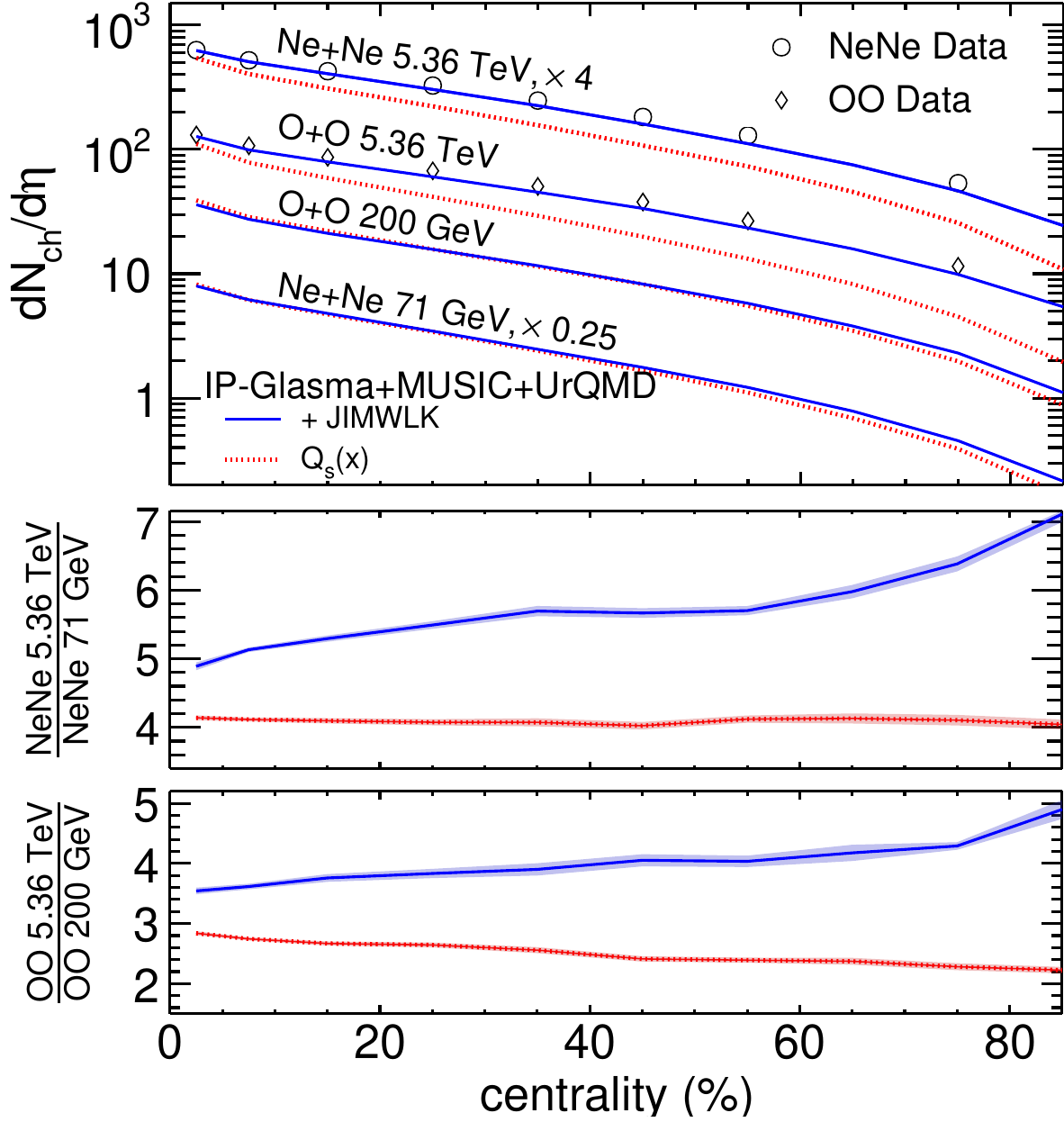}
    \caption{Model predictions for the charged hadron multiplicity distributions in O+O and Ne+Ne collisions at the LHC  5.36 TeV and at RHIC 200 GeV and 71 GeV energies. The preliminary ALICE data is from~\cite{ALICE:MartaUrioni_IS2025}}
    \label{fig:dNch_OO}
\end{figure}

Oxygen-oxygen collisions provide a unique system to probe the small-$x$ evolution effects in the initial state, because they have been measured at both RHIC at $\sqrt{s}=200$ GeV and LHC at $\sqrt{s}=5.36$ TeV~\cite{Brewer:2021kiv,ALICE:2025luc,ATLAS:2025nnt,cms:OOnote,Yan:2025kwz}. Furthermore, because of the smaller size of the oxygen nuclei compared to lead, initial state fluctuations (and their possible modification due to JIMWLK evolution) can be expected to have a more pronounced effect on flow observables. 
The LHC also measured Ne+Ne collisions at the same energy as O+O, which is particularly interesting because of the expected differences in shape of these two nuclei \cite{Giacalone:2025vxa, cms:OOnote,ALICE:2025luc,ATLAS:2025nnt}. The energy dependence in the Ne+Ne system could further be explored by comparing fixed target Ne+Ne collisions (using SMOGII~\cite{lhcb_smog} at LHCb) with Ne+Ne collisions in collider mode.

We begin the analysis of O+O and Ne+Ne collisions by showing in Fig.~\ref{fig:dNch_OO} the midrapidity charged particle multiplicity as a function of centrality at different $\sqrt{s}$ probed at RHIC and the LHC. As already seen in the case of Pb+Pb collisions, the small-$x$ evolution in the nuclear geometry has a significant effect on the multiplicity distribution. For these smaller collision systems, the differences between the two initial state descriptions become visible at even smaller centralities than for Pb+Pb collisions. For all collision systems and center-of-mass energies, the JIMWLK evolution results in flatter distributions compared to the $ Q_s(x)$ setup. 

The multiplicity ratios of identical collision systems at different energies shown in the lower panel of Fig.~\ref{fig:dNch_OO} are predictions that can be easily tested experimentally. With the JIMWLK evolution included, these ratios increase significantly with centrality.
Observing this trend experimentally could provide evidence for the presence of effects of JIMWLK evolution in the multiplicity distributions. 

\begin{figure}[tb]
    \centering
    \includegraphics[width=\columnwidth]{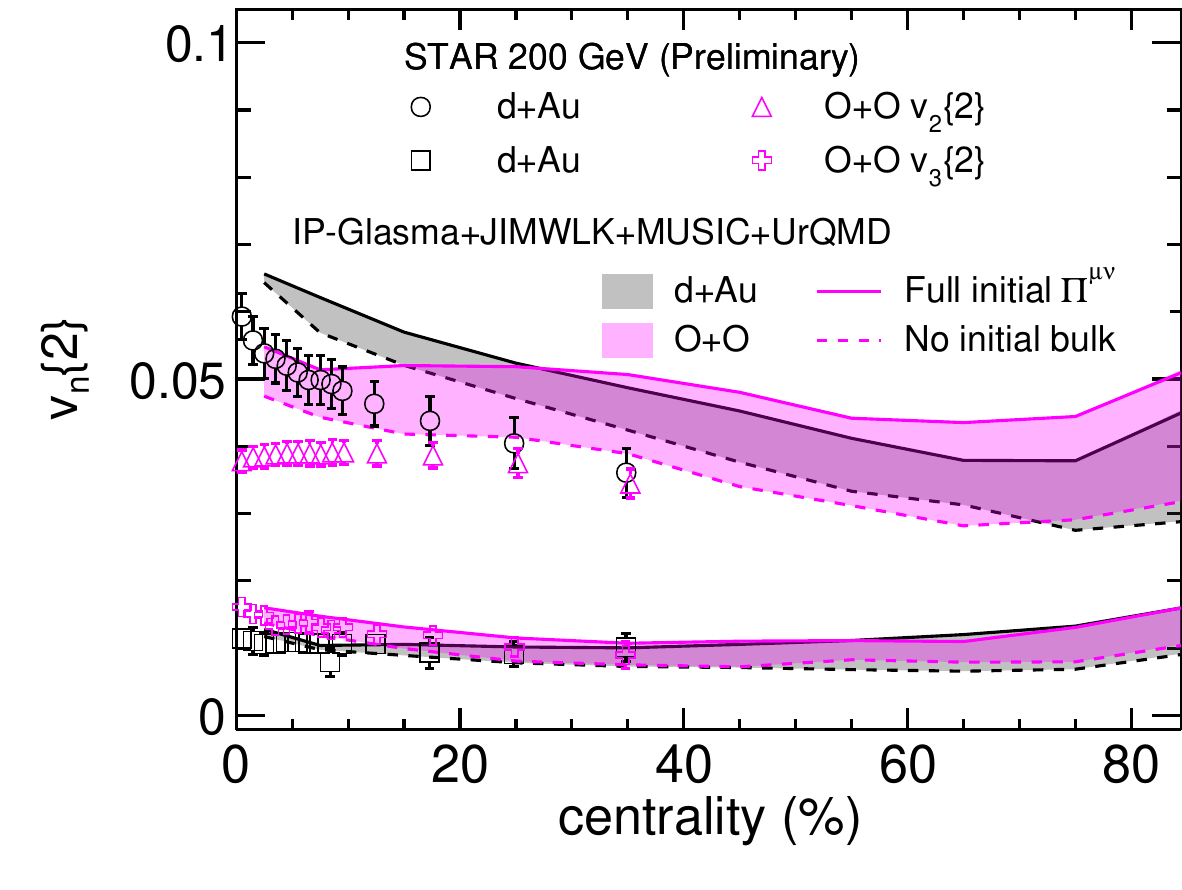}
    \includegraphics[width=\columnwidth]{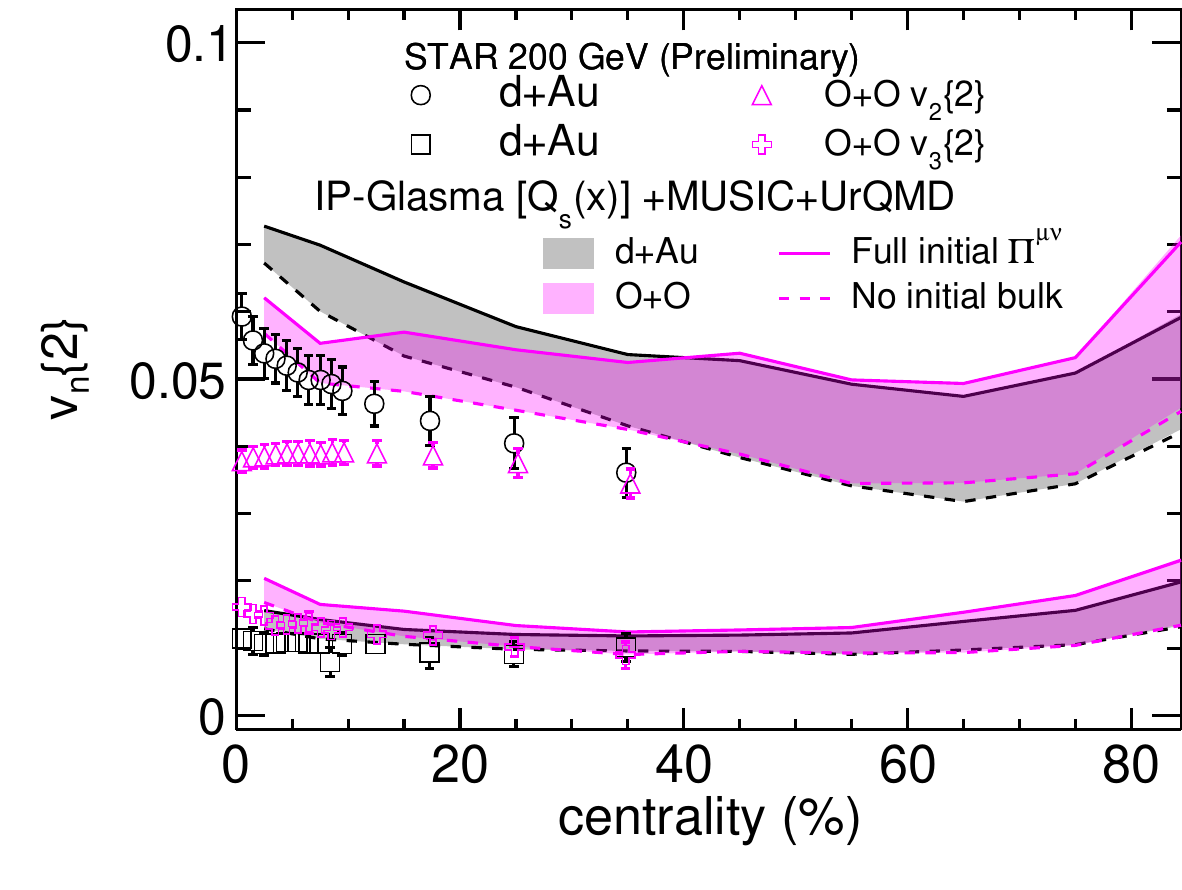}
    \caption{Collective flow $v_n$ in O+O and d+Au collisions at $\sqrt{s}=200$ GeV calculated with the ``+JIMWLK'' (upper panel) and ``$Q_s(x)$'' (lower panel) setups. The preliminary STAR data are from \cite{Yan:2025kwz,STAR:2025ivi}. }
    \label{fig:vn_OO_STAR}
\end{figure}

\begin{figure}[tb]
    \centering
    \includegraphics[width=\columnwidth]{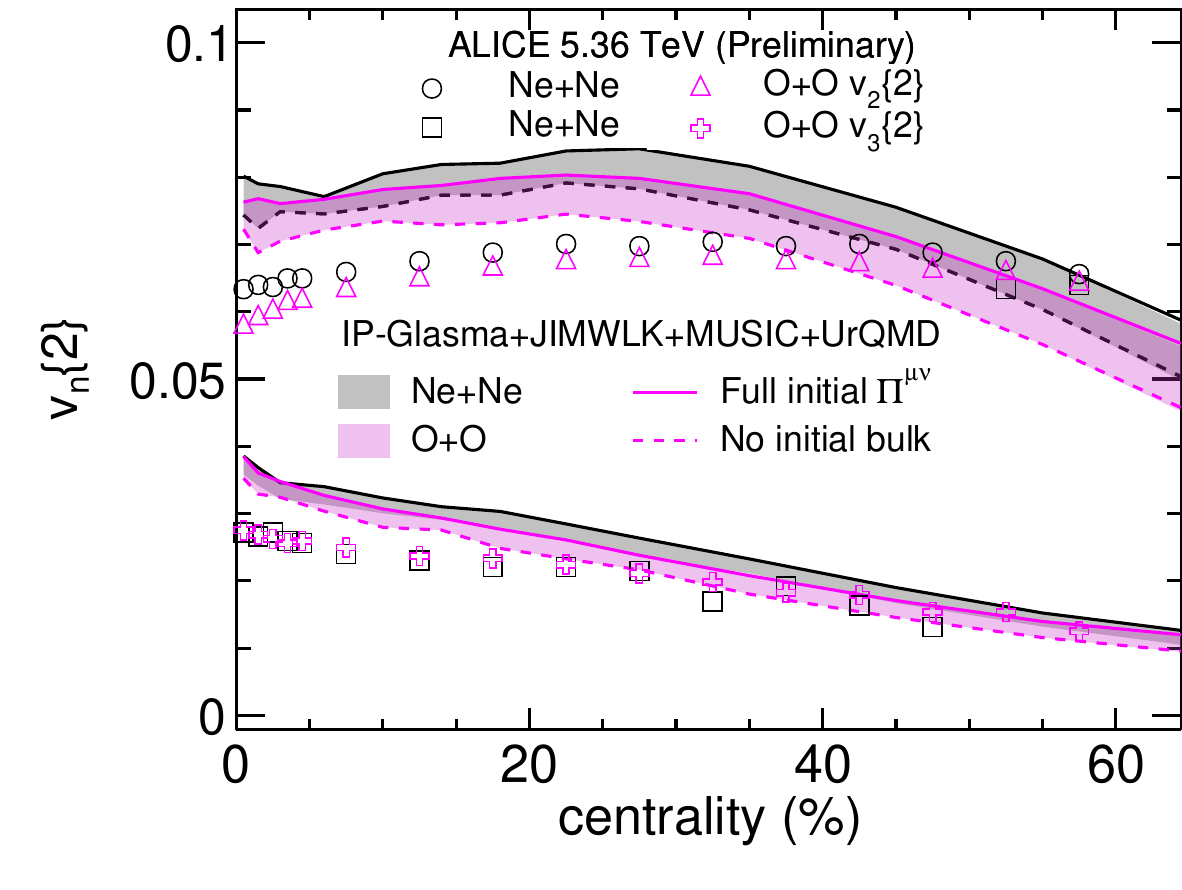}
    \includegraphics[width=\columnwidth]{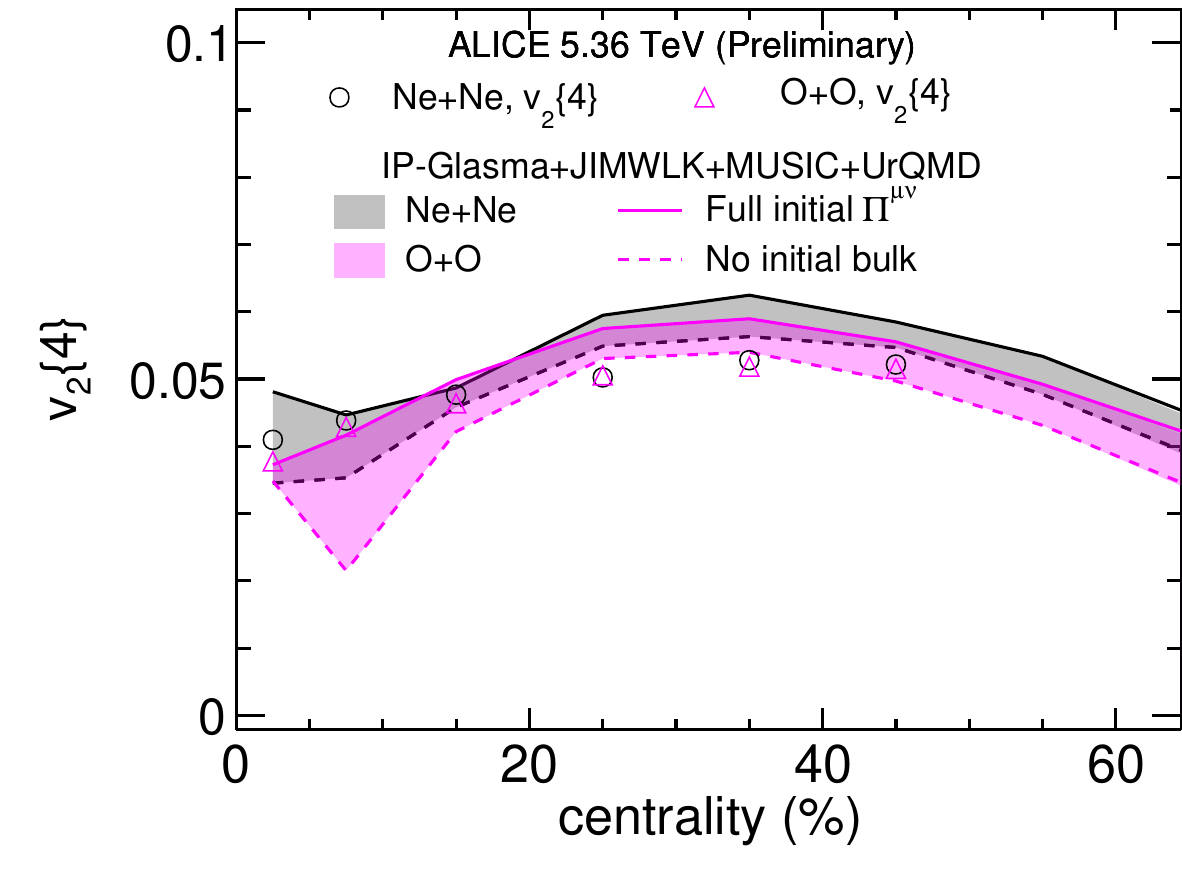}
    \caption{Model predictions for the collective flow $v_n\{2\}$, $n = 2, 3$ (upper panel) and $v_2\{4\}$ (lower panel) in O+O and Ne+Ne collisions at 5.36 TeV. The preliminary ALICE data are from \cite{ALICE:2025luc}. 
    }
    \label{fig:vn_OONeNe}
\end{figure}

\begin{figure}[tb]
    \centering
    \includegraphics[width=\columnwidth]{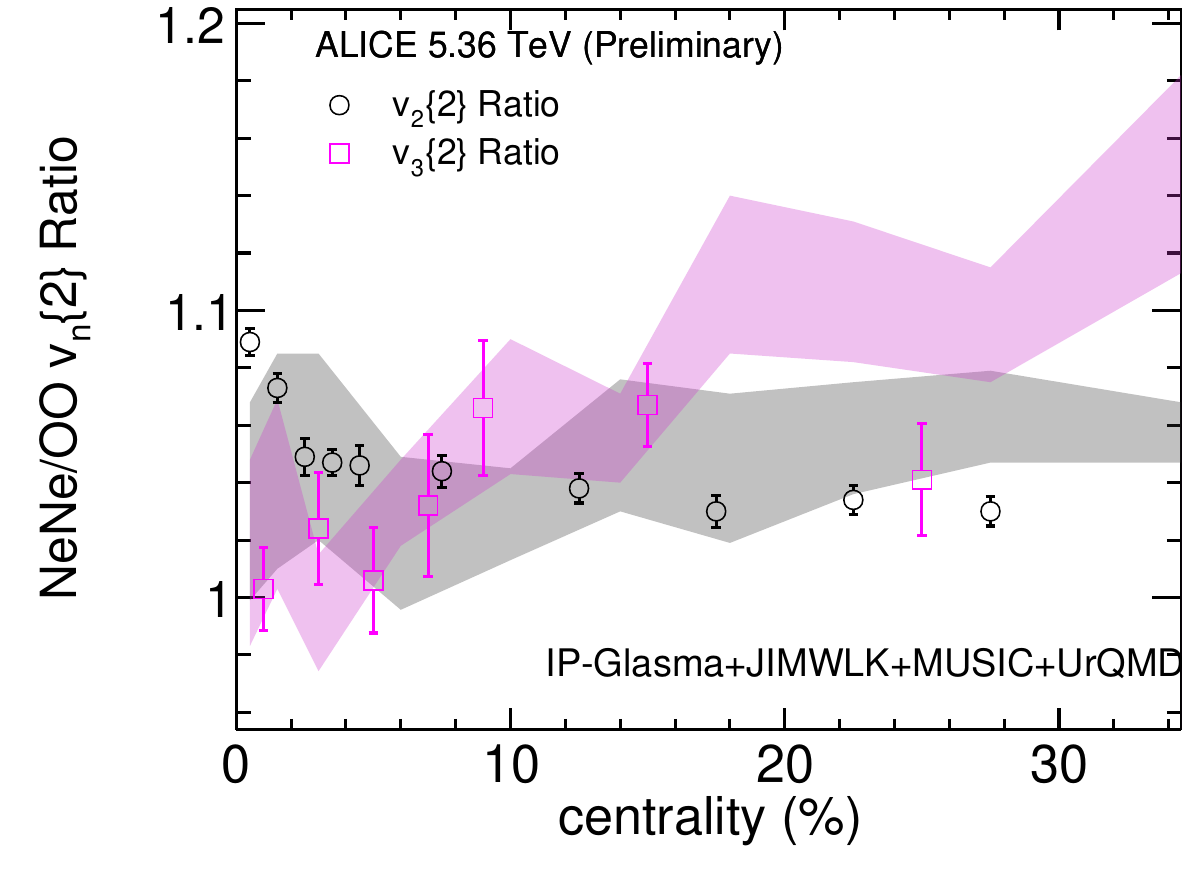}
    \caption{Model predictions for the collective flow $v_n\{2\}$ ratio ($n = 2, 3$) between O+O and Ne+Ne collisions at 5.36 TeV.  The preliminary ALICE data are from \cite{ALICE:2025luc}.}
    \label{fig:vnRatio_OONeNe}
\end{figure}

\begin{figure}[tb]
    \centering
    \includegraphics[width=\columnwidth]{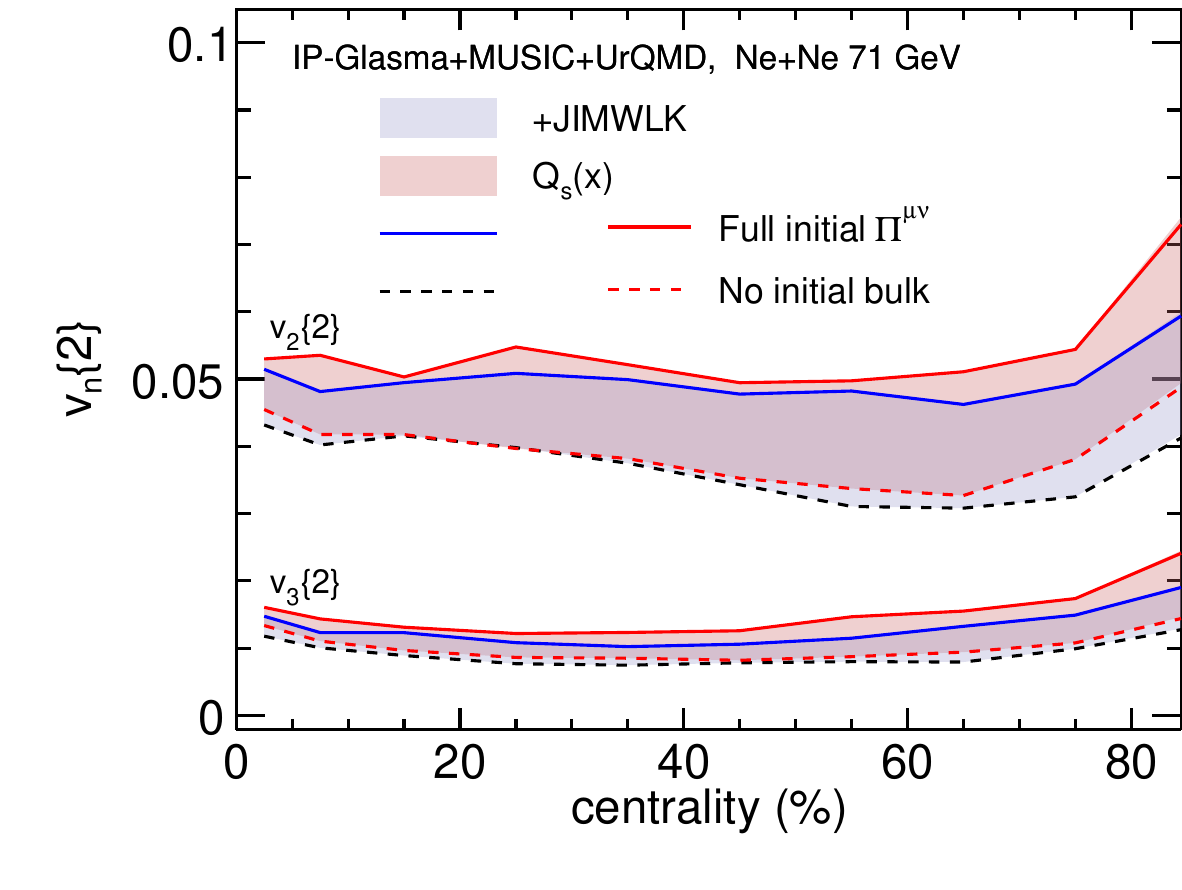}
    \caption{Model predictions for the collective flow $v_n$ with ``+JIMWLK" and ``$Q_s(x)$" setups in Ne+Ne collisions at 71 GeV.}
    \label{fig:vn_NeNe71}
\end{figure}

Comparisons to the RHIC elliptic and triangular flow measurements in Oxygen-Oxygen collisions are shown in Fig.~\ref{fig:vn_OO_STAR}. 
As a reference, the corresponding d+Au measurements at the same center-of-mass energy are also shown. For both initial state prescriptions, two sets of results are shown as upper and lower ends of a band: We either initialize with the full shear-stress tensor $\Pi^{\mu\nu}$ of the energy-momentum tensor extracted from the CYM evolution (our standard method), or set the bulk stress to zero. The bulk stress term does not reflect a real bulk viscosity, but it absorbs differences between the conformal equation of state (EOS) in the Glasma and the QCD EOS in the hydrodynamic medium~\cite{Schenke:2020mbo}. Thus, the band reflects the systematic uncertainty from matching the conformal glasma initial state to the near-equilibrium QCD fluid. We did not show this uncertainty band for observables in heavy ion (e.g.~Au, Pb) collisions, because in such large systems the uncertainty from varying the initial viscous tensor is much smaller. 

Comparing the two panels in Fig.~\ref{fig:vn_OO_STAR}, we find that with the JIMWLK evolution in the initial state the flow harmonics are suppressed compared to the $Q_s(x)$ setup at all centralities and for both O+O and d+Au collisions. A slightly better agreement with the RHIC data is obtained when the ``+JIMWLK'' setup is used, although we overestimate the elliptic flow for all centralities. The qualitative differences between d+Au and O+O flow harmonics are well reproduced, including the fact that they are similar in more peripheral collisions starting from $\sim 30\%$ centrality.

Predictions for the high-energy O+O and Ne+Ne collisions measured at the LHC are shown in Figs.~\ref{fig:vn_OONeNe} and \ref{fig:vnRatio_OONeNe}.
We observe increased $v_2\{2\}$ for the Ne+Ne system, a result expected from the more strongly deformed shape of the neon nucleus. The difference between the $v_3\{2\}$ values is smaller.
Our calculations of anisotropic flows overestimate the ALICE preliminary data for $v_n\{2\}$ but are comparable for $v_2\{4\}$. Because $v_n\{2\}$ and $v_n\{4\}$ receive opposite contributions from flow fluctuations, the current comparisons in the central collisions suggest that the model contains more flow fluctuations than those in the measurements. The centrality dependence of the $v_n\{2\}$ ratios between the two small systems is reproduced by our model, while the current theoretical uncertainty from the pre-equilibrium phase is significant.

Predictions for the elliptic and triangular flow in Ne+Ne collisions at $\sqrt{s}=71$ GeV are shown in Fig.~\ref{fig:vn_NeNe71}. In this kinematics, there is only very little evolution from $x=0.01$ to $x=0.0067$. Nevertheless, we observe a slight reduction of both $v_2\{2\}$ and $v_3\{2\}$ when the JIMWLK evolution is included. Similarly to the case of 200 GeV collisions shown in Fig.~\ref{fig:vn_OO_STAR}, results show relatively large sensitivity to the description of the initial $\Pi^{\mu\nu}$. Given this uncertainty, the effect of JIMWLK evolution can not be clearly distinguished. 

\subsection{p+Pb collisions}
Finally, we study an even smaller collision system, proton-lead collisions at $\sqrt{s}=5.02$ TeV measured at the LHC. In such small systems, the sensitivity to the nucleon geometry is maximal, as the proton event-by-event structure dominates the initial state eccentricities. As already discussed in Ref.~\cite{Mantysaari:2017cni}, proton substructure fluctuations constrained by HERA data are crucial to understand flow harmonics measured in p+Pb collisions (see also an earlier analysis of Ref.~\cite{Schenke:2014zha}). In the companion letter~\cite{Mantysaari:2025tcg} we demonstrated that the multiplicity distributions in such small systems are sensitive to the high-energy evolution in the initial state, and that the ALICE and PHENIX data~\cite{PHENIX:2003iij,ALICE:2013wgn} are better described by the ``+JIMWLK'' setup. In this work, we focus on flow observables.

\begin{figure}[tb]
    \centering
    \includegraphics[width=\columnwidth]{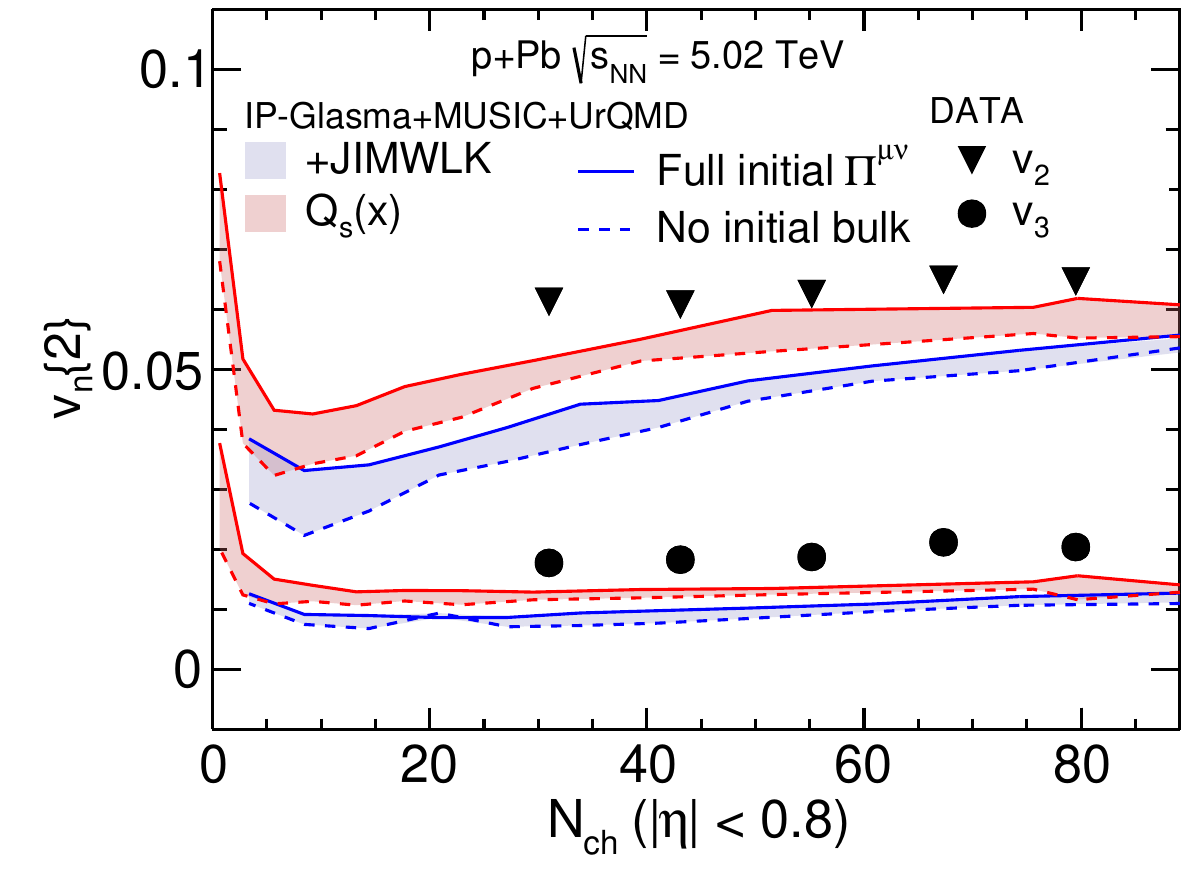}
    \caption{Collective flow $v_n$ with ``+JIMWLK" and ``$Q_s(x)$" setups in p+Pb collisions at the LHC 5.02 TeV. Experimental data is from the ALICE Collaboration \cite{ALICE:2019zfl}.}
    \label{fig:vn_pPb}
\end{figure}

The flow harmonics in p+Pb collisions are shown in Fig.~\ref{fig:vn_pPb}. We find a larger effect of the inclusion of JIMWLK evolution on $v_n\{2\}$ coefficients compared to Pb+Pb collisions shown in Fig.~\ref{fig:PbPb_vn}. 
The JIMWLK evolution again decreases the $v_2\{2\}$ and $v_3\{2\}$ flow harmonics. Unlike in other collision systems studied in this work, the available ALICE data~\cite{ALICE:2019zfl} for proton-nucleus collisions favor the $Q_s(x)$ setup. 
This indicates a preference for stronger subnucleonic fluctuations than those present in the +JIMWLK setup after the evolution. We note, however, that the proton-nucleus collisions are the most challenging systems to describe within a hydrodynamical framework~\cite{Gallmeister:2018mcn, Schenke:2020mbo, Noronha:2024dtq}, and a fully three-dimensional setup may be required to accurately describe correlation measurements involving a rapidity gap in a system where boost invariance is not as good an approximation as in heavy ion collisions.

\begin{figure}[tb]
    \centering
    \includegraphics[width=\columnwidth]{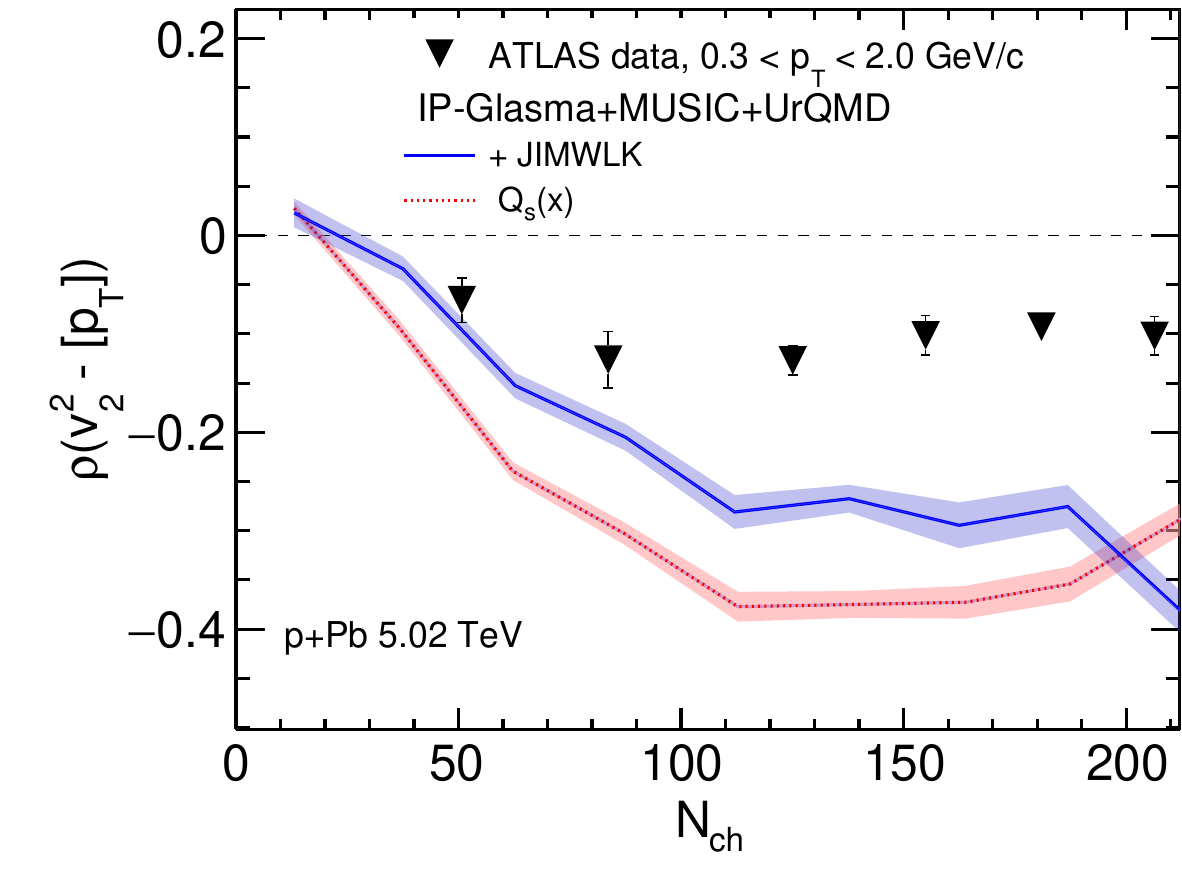}
    \caption{The $v_2^2-p_T$ correlator in p+Pb collisions at the LHC 5.02 TeV. The ATLAS data is
 from \cite{ATLAS:2019pvn}}
    \label{fig:rho_pPb}
\end{figure}

Finally, we calculate the correlation between the squared elliptic flow $v_2^2$ and the mean transverse momentum in p+Pb collisions (results for the mean $p_T$ were shown in \cite{Mantysaari:2025tcg}), and present the results in Fig.~\ref{fig:rho_pPb}. This correlation is found to depend strongly on the initial-state description. In particular, including the JIMWLK evolution decreases the magnitude of the correlation by $\sim 40\%$ in the central $N_\mathrm{ch}\gtrsim 100$ collisions.  This sensitivity arises because both the elliptic flow and $\langle p_T\rangle$ are directly sensitive to the lumpiness of the nucleon density profile in p+Pb collisions. While including the JIMWLK evolution results in a better description of the ATLAS data~\cite{ATLAS:2019pvn}, the strength of this correlator is still significantly overestimated except in the lowest $N_\mathrm{ch}$ bin.

In this paper, We simulated 5K minimum-bias events for Pb–Pb and Au–Au collisions,  40K minimum-bias events for p–Pb collision and Ne-Ne at 72 GeV, and 160K for Ne-Ne and O-O at 5.36 TeV. In several figures (Figures. 6-10), the error bands shown represent the systematic uncertainties from matching the Glasma stage to hydrodynamics. The statistical errors are much smaller than these systematic errors and are not explicitly shown.

\section{Conclusions}
We have implemented a QCD based treatment of the collision–energy dependence of the initial state by coupling JIMWLK evolution to the IP-Glasma framework. The resulting energy-momentum tensor is propagated through viscous hydrodynamics, which in turn is coupled to a hadronic afterburner that evolves the low density regime. Constraining the non-perturbative parameters with exclusive $\gamma{+}p \to \mathrm{J}/\psi{+}p$ data and fixing a single overall normalization and transport coefficients to Au+Au at $\sqrt{s_{\rm NN}}{=}200$~GeV, we obtained bona fide predictions across systems and energies without retuning.

The inclusion of JIMWLK evolution modifies charged hadron multiplicities in a characteristic way. The growth and smoothing of nuclear gluon density profiles at smaller $x$ produce flatter centrality trends in A+A collisions at high energy, yielding an increasing centrality dependence of the multiplicity ratios between LHC and RHIC energies. 

Further, anisotropic flow observables are systematically affected. Relative to the baseline with only $Q_s(x)$ dependence, the ``+JIMWLK'' initial state reduces $v_{2,3,4}\{2\}$ in both Au+Au ($200$~GeV) and Pb+Pb ($5.02$~TeV). Further, the approximately flat $v_2\{4\}/v_2\{2\}$ ratio obtained with JIMWLK is closer to the ALICE measurements, indicating suppressed initial-state fluctuations due to the evolution-driven smoothing. The same mechanism lowers $\langle p_T\rangle$ at the LHC compared to a static geometry and improves agreement with data.

The sensitivity to small-$x$ dynamics is amplified in smaller systems. In O+O and Ne+Ne collisions, we predicted noticeably flatter centrality dependence of the charged hadron multiplicity with JIMWLK and a clear system ordering of flow harmonics (larger $v_2$ in Ne+Ne than O+O, consistent with stronger intrinsic deformation). Preliminary experimental data from the LHC show excellent agreement with our calculations of the ratio between flow harmonics in Ne+Ne vs.~O+O, while the absolute $v_n\{2\}$ values are overestimated for small centralities in both systems.

In p+Pb at $5.02$~TeV, evolution reduces $v_{2,3}\{2\}$ and the magnitude of the $v_2^2$–$\langle p_T\rangle$ correlator, underscoring the role of nucleon substructure energy evolution when the proton drives the geometry.

Our analysis highlights that omitting energy-dependent geometry evolution can bias the extraction of QGP transport coefficients in global fits. While our midrapidity, boost-invariant setup already reveals clear trends, several controlled extensions are well motivated: (i) fully 3+1D simulations to simultaneously capture rapidity and $\sqrt{s}$ dependencies; (ii) a broader and more statistics hungry set of observables, including e.g.~symmetric cumulants, event-plane correlators, rapidity decorrelations, and mean-$p_T$ fluctuations; (iii) a complete global Bayesian analysis to constrain parameters using data from a wide range of systems and collision energies. 

The data that supports the findings of this article are openly available \cite{data_link}.

\begin{acknowledgments}
This material is based upon work supported by the U.S. Department of Energy, Office of Science, Office of Nuclear Physics, under DOE Contract No.~DE-SC0012704 (B.P.S.) and Award No.~DE-SC0021969 (C.S.), and within the framework of the Saturated Glue (SURGE) Topical Theory Collaboration.
C.S. acknowledges a DOE Office of Science Early Career Award. 
H.M. is supported by the Research Council of Finland, the Centre of Excellence in Quark Matter, and projects 338263 and 359902, and by the European Research Council (ERC, grant agreements No. ERC-2023-101123801 GlueSatLight and ERC-2018-ADG-835105 YoctoLHC).
W.B.Z. was supported by DOE under Contract No. DE-AC02-05CH11231, by NSF under Grant No. OAC-2004571 within the X-SCAPE Collaboration, and within the framework of the SURGE Topical Theory Collaboration. 
The content of this article does not reflect the official opinion of the European Union and responsibility for the information and views expressed therein lies entirely with the authors.
This research was done using computational resources provided by the Open Science Grid (OSG)~\cite{Pordes:2007zzb,Sfiligoi:2009cct,OSPool,OSDF}, which is supported by the National Science Foundation awards \#2030508 and \#2323298.
Computing resources from CSC – IT Center for Science in Espoo, Finland and the Finnish Grid and Cloud Infrastructure (persistent identifier \texttt{urn:nbn:fi:research-infras-2016072533}) were also used in this work.
\end{acknowledgments}

\bibliographystyle{JHEP-2modlong.bst}
\bibliography{refs}

@article{Mantysaari:2022ffw,
    author = {M\"antysaari, Heikki and Schenke, Bj\"orn and Shen, Chun and Zhao, Wenbin},
    title = "{Bayesian inference of the fluctuating proton shape}",
    eprint = "2202.01998",
    archivePrefix = "arXiv",
    primaryClass = "hep-ph",
    doi = "10.1016/j.physletb.2022.137348",
    journal = "Phys. Lett. B",
    volume = "833",
    pages = "137348",
    year = "2022"
}

@article{Giacalone:2024ixe,
    author = "Giacalone, Giuliano and others",
    title = "{Anisotropic Flow in Fixed-Target $^{208}$Pb+$^{20}$Ne Collisions as a Probe of Quark-Gluon Plasma}",
    eprint = "2405.20210",
    archivePrefix = "arXiv",
    primaryClass = "nucl-th",
    reportNumber = "CERN-TH-2024-074",
    doi = "10.1103/PhysRevLett.134.082301",
    journal = "Phys. Rev. Lett.",
    volume = "134",
    number = "8",
    pages = "082301",
    year = "2025"
}

@article{ALICE:2025luc,
    author = "Abualrob, Ibrahim Jaser and others",
    collaboration = "ALICE",
    title = "{Evidence of nuclear geometry-driven anisotropic flow in OO and Ne$-$Ne collisions at $\mathbf{\sqrt{{\textit s}_{\rm\mathbf {NN}}}}$ = 5.36 TeV}",
    eprint = "2509.06428",
    archivePrefix = "arXiv",
    primaryClass = "nucl-ex",
    reportNumber = "CERN-EP-2025-203",
    month = "9",
    year = "2025"
}

@article{Voloshin:2008dg,
    author = "Voloshin, Sergei A. and Poskanzer, Arthur M. and Snellings, Raimond",
    editor = "Stock, R.",
    title = "{Collective phenomena in non-central nuclear collisions}",
    eprint = "0809.2949",
    archivePrefix = "arXiv",
    primaryClass = "nucl-ex",
    doi = "10.1007/978-3-642-01539-7_10",
    journal = "Landolt-Bornstein",
    volume = "23",
    pages = "293--333",
    year = "2010"
}

@article{Gallmeister:2018mcn,
    author = "Gallmeister, K. and Niemi, H. and Greiner, C. and Rischke, D. H.",
    title = "{Exploring the applicability of dissipative fluid dynamics to small systems by comparison to the Boltzmann equation}",
    eprint = "1804.09512",
    archivePrefix = "arXiv",
    primaryClass = "nucl-th",
    doi = "10.1103/PhysRevC.98.024912",
    journal = "Phys. Rev. C",
    volume = "98",
    number = "2",
    pages = "024912",
    year = "2018"
}

@article{ATLAS:2025nnt,
    author = "Aad, Georges and others",
    collaboration = "ATLAS",
    title = "{Measurement of the azimuthal anisotropy of charged particles in $\sqrt{s_{\mathrm{NN}}}=5.36$ TeV $^{16}$O$+^{16}$O and $^{20}$Ne$+^{20}$Ne collisions with the ATLAS detector}",
    eprint = "2509.05171",
    archivePrefix = "arXiv",
    primaryClass = "nucl-ex",
    reportNumber = "CERN-EP-2025-200",
    month = "9",
    year = "2025"
}

@article{Giacalone:2024luz,
    author = "Giacalone, Giuliano and others",
    title = "{Exploiting $^{20}$Ne Isotopes for Precision Characterizations of Collectivity in Small Systems}",
    eprint = "2402.05995",
    archivePrefix = "arXiv",
    primaryClass = "nucl-th",
    reportNumber = "CERN-TH-2024-021",
    doi = "10.1103/k8rb-jgvq",
    journal = "Phys. Rev. Lett.",
    volume = "135",
    number = "1",
    pages = "012302",
    year = "2025"
}

@article{Noronha:2024dtq,
    author = {Noronha, Jorge and Schenke, Bj{\"o}rn and Shen, Chun and Zhao, Wenbin},
    title = "{Progress and challenges in small systems}",
    eprint = "2401.09208",
    archivePrefix = "arXiv",
    primaryClass = "nucl-th",
    doi = "10.1142/9789811294679_0004",
    journal = "Int. J. Mod. Phys. E",
    volume = "33",
    number = "06",
    pages = "2430005",
    year = "2024"
}

@article{Kowalski:2003hm,
    author = "Kowalski, Henri and Teaney, Derek",
    title = "{An Impact parameter dipole saturation model}",
    eprint = "hep-ph/0304189",
    archivePrefix = "arXiv",
    doi = "10.1103/PhysRevD.68.114005",
    journal = "Phys. Rev. D",
    volume = "68",
    pages = "114005",
    year = "2003"
}

@article{ALICE:2015juo,
    author = "Adam, Jaroslav and others",
    collaboration = "ALICE",
    title = "{Centrality Dependence of the Charged-Particle Multiplicity Density at Midrapidity in Pb-Pb Collisions at $\sqrt{s_{\rm NN}}$ = 5.02 TeV}",
    eprint = "1512.06104",
    archivePrefix = "arXiv",
    primaryClass = "nucl-ex",
    reportNumber = "CERN-PH-EP-2015-324, ALICE-PUBLIC-2015-008",
    doi = "10.1103/PhysRevLett.116.222302",
    journal = "Phys. Rev. Lett.",
    volume = "116",
    number = "22",
    pages = "222302",
    year = "2016"
}

@article{H1:2005dtp,
    author = "Aktas, A. and others",
    collaboration = "H1",
    title = "{Elastic $\mathrm{J}/\psi$ production at HERA}",
    eprint = "hep-ex/0510016",
    archivePrefix = "arXiv",
    reportNumber = "DESY-05-161",
    doi = "10.1140/epjc/s2006-02519-5",
    journal = "Eur. Phys. J. C",
    volume = "46",
    pages = "585--603",
    year = "2006"
}

@article{ZEUS:2002wfj,
    author = "Chekanov, S. and others",
    collaboration = "ZEUS",
    title = "{Exclusive photoproduction of $\mathrm{J}/\psi$ mesons at HERA}",
    eprint = "hep-ex/0201043",
    archivePrefix = "arXiv",
    reportNumber = "DESY-02-008",
    doi = "10.1007/s10052-002-0953-7",
    journal = "Eur. Phys. J. C",
    volume = "24",
    pages = "345--360",
    year = "2002"
}

@article{ALICE:2014eof,
    author = "Abelev, Betty Bezverkhny and others",
    collaboration = "ALICE",
    title = "{Exclusive $\mathrm{J/}\psi$ photoproduction off protons in ultra-peripheral p-Pb collisions at $\sqrt{s_{\rm NN}}=5.02$ TeV}",
    eprint = "1406.7819",
    archivePrefix = "arXiv",
    primaryClass = "nucl-ex",
    reportNumber = "CERN-PH-EP-2014-149",
    doi = "10.1103/PhysRevLett.113.232504",
    journal = "Phys. Rev. Lett.",
    volume = "113",
    number = "23",
    pages = "232504",
    year = "2014"
}

@article{ALICE:2018oyo,
    author = "Acharya, Shreyasi and others",
    collaboration = "ALICE",
    title = "{Energy dependence of exclusive $\mathrm {J}/\psi $ photoproduction off protons in ultra-peripheral p\textendash{}Pb collisions at $\sqrt{s_{\mathrm {\scriptscriptstyle NN}}} = 5.02$ TeV}",
    eprint = "1809.03235",
    archivePrefix = "arXiv",
    primaryClass = "nucl-ex",
    reportNumber = "CERN-EP-2018-236",
    doi = "10.1140/epjc/s10052-019-6816-2",
    journal = "Eur. Phys. J. C",
    volume = "79",
    number = "5",
    pages = "402",
    year = "2019"
}

@article{Bertulani:2005ru,
    author = "Bertulani, Carlos A. and Klein, Spencer R. and Nystrand, Joakim",
    title = "{Physics of ultra-peripheral nuclear collisions}",
    eprint = "nucl-ex/0502005",
    archivePrefix = "arXiv",
    doi = "10.1146/annurev.nucl.55.090704.151526",
    journal = "Ann. Rev. Nucl. Part. Sci.",
    volume = "55",
    pages = "271--310",
    year = "2005"
}

@article{Klein:2019qfb,
    author = {Klein, Spencer R. and M\"antysaari, Heikki},
    title = "{Imaging the nucleus with high-energy photons}",
    eprint = "1910.10858",
    archivePrefix = "arXiv",
    primaryClass = "hep-ex",
    doi = "10.1038/s42254-019-0107-6",
    journal = "Nature Rev. Phys.",
    volume = "1",
    number = "11",
    pages = "662--674",
    year = "2019"
}

@article{LHCb:2014acg,
    author = "Aaij, Roel and others",
    collaboration = "LHCb",
    title = "{Updated measurements of exclusive $J/\psi$ and $\psi$(2S) production cross-sections in pp collisions at $\sqrt{s}=7$ TeV}",
    eprint = "1401.3288",
    archivePrefix = "arXiv",
    primaryClass = "hep-ex",
    reportNumber = "CERN-PH-EP-2013-233, LHCB-PAPER-2013-059",
    doi = "10.1088/0954-3899/41/5/055002",
    journal = "J. Phys. G",
    volume = "41",
    pages = "055002",
    year = "2014"
}

@article{LHCb:2018rcm,
    author = "Aaij, Roel and others",
    collaboration = "LHCb",
    title = "{Central exclusive production of $J/\psi$ and $\psi(2S)$ mesons in $pp$ collisions at $\sqrt{s}=13~$TeV}",
    eprint = "1806.04079",
    archivePrefix = "arXiv",
    primaryClass = "hep-ex",
    reportNumber = "LHCB-PAPER-2018-011, LHCb-PAPER-2018-011, CERN-EP-2018-152",
    doi = "10.1007/JHEP10(2018)167",
    journal = "JHEP",
    volume = "10",
    pages = "167",
    year = "2018"
}

@inproceedings{Heinz:2024jwu,
    author = {Heinz, Ulrich and Schenke, Bj\"orn},
    title = "{Hydrodynamic Description of the Quark-Gluon Plasma}",
    eprint = "2412.19393",
    archivePrefix = "arXiv",
    primaryClass = "nucl-th",
    month = "12",
    year = "2024"
}

@article{Jalilian-Marian:1997qno,
    author = "Jalilian-Marian, Jamal and Kovner, Alex and Leonidov, Andrei and Weigert, Heribert",
    title = "{The BFKL equation from the Wilson renormalization group}",
    eprint = "hep-ph/9701284",
    archivePrefix = "arXiv",
    reportNumber = "TPI-MINN-96-28-T, NUC-MINN-96-22-T, HEP-MINN-96-1524",
    doi = "10.1016/S0550-3213(97)00440-9",
    journal = "Nucl. Phys. B",
    volume = "504",
    pages = "415--431",
    year = "1997"
}

@article{ALICE:2013wgn,
    author = "Abelev, Betty Bezverkhny and others",
    collaboration = "ALICE",
    title = "{Multiplicity Dependence of Pion, Kaon, Proton and Lambda Production in p-Pb Collisions at $\sqrt{s_{NN}}$ = 5.02 TeV}",
    eprint = "1307.6796",
    archivePrefix = "arXiv",
    primaryClass = "nucl-ex",
    reportNumber = "CERN-PH-EP-2013-135",
    doi = "10.1016/j.physletb.2013.11.020",
    journal = "Phys. Lett. B",
    volume = "728",
    pages = "25--38",
    year = "2014"
}

@article{PHENIX:2003iij,
    author = "Adler, S. S. and others",
    collaboration = "PHENIX",
    title = "{Identified charged particle spectra and yields in Au+Au collisions at $\sqrt{s_{NN}}=200$ GeV}",
    eprint = "nucl-ex/0307022",
    archivePrefix = "arXiv",
    doi = "10.1103/PhysRevC.69.034909",
    journal = "Phys. Rev. C",
    volume = "69",
    pages = "034909",
    year = "2004"
}

@article{ALICE:2018hza,
    author = "Acharya, Shreyasi and others",
    collaboration = "ALICE",
    title = "{Transverse momentum spectra and nuclear modification factors of charged particles in Xe-Xe collisions at $\sqrt{s_{\rm NN}}$ = 5.44 TeV}",
    eprint = "1805.04399",
    archivePrefix = "arXiv",
    primaryClass = "nucl-ex",
    reportNumber = "CERN-EP-2018-112",
    doi = "10.1016/j.physletb.2018.10.052",
    journal = "Phys. Lett. B",
    volume = "788",
    pages = "166--179",
    year = "2019"
}

@article{H1:2013okq,
    author = "Alexa, C. and others",
    collaboration = "H1",
    title = "{Elastic and Proton-Dissociative Photoproduction of J/psi Mesons at HERA}",
    eprint = "1304.5162",
    archivePrefix = "arXiv",
    primaryClass = "hep-ex",
    reportNumber = "DESY-13-058",
    doi = "10.1140/epjc/s10052-013-2466-y",
    journal = "Eur. Phys. J. C",
    volume = "73",
    number = "6",
    pages = "2466",
    year = "2013"
}

@article{Mantysaari:2022sux,
    author = {M\"antysaari, Heikki and Salazar, Farid and Schenke, Bj\"orn},
    title = "{Nuclear geometry at high energy from exclusive vector meson production}",
    eprint = "2207.03712",
    archivePrefix = "arXiv",
    primaryClass = "hep-ph",
    doi = "10.1103/PhysRevD.106.074019",
    journal = "Phys. Rev. D",
    volume = "106",
    number = "7",
    pages = "074019",
    year = "2022"
}

@article{Garcia-Montero:2025hys,
    author = {Garcia-Montero, Oscar and Schlichting, S{\"o}ren},
    title = "{Effective theories for nuclei at high energies}",
    eprint = "2502.09721",
    archivePrefix = "arXiv",
    primaryClass = "hep-ph",
    doi = "10.1140/epja/s10050-025-01523-7",
    journal = "Eur. Phys. J. A",
    volume = "61",
    number = "3",
    pages = "54",
    year = "2025"
}

@article{Kuha:2024kmq,
    author = "Kuha, Mikko and Auvinen, Jussi and Eskola, Kari J. and Hirvonen, Henry and Kanakubo, Yuuka and Niemi, Harri",
    title = "{Monte Carlo event generator MC-EKRT with saturated minijet production for initializing (3+1)D fluid dynamics in high-energy nuclear collisions}",
    eprint = "2406.17592",
    archivePrefix = "arXiv",
    primaryClass = "hep-ph",
    doi = "10.1103/PhysRevC.111.054914",
    journal = "Phys. Rev. C",
    volume = "111",
    number = "5",
    pages = "054914",
    year = "2025"
}

@article{Mantysaari:2020axf,
    author = {M\"antysaari, Heikki},
    title = "{Review of proton and nuclear shape fluctuations at high energy}",
    eprint = "2001.10705",
    archivePrefix = "arXiv",
    primaryClass = "hep-ph",
    doi = "10.1088/1361-6633/aba347",
    journal = "Rept. Prog. Phys.",
    volume = "83",
    number = "8",
    pages = "082201",
    year = "2020"
}

@article{Mantysaari:2017cni,
    author = {M\"antysaari, Heikki and Schenke, Bj\"orn and Shen, Chun and Tribedy, Prithwish},
    title = "{Imprints of fluctuating proton shapes on flow in proton-lead collisions at the LHC}",
    eprint = "1705.03177",
    archivePrefix = "arXiv",
    primaryClass = "nucl-th",
    doi = "10.1016/j.physletb.2017.07.038",
    journal = "Phys. Lett. B",
    volume = "772",
    pages = "681--686",
    year = "2017"
}

@article{Schenke:2014zha,
    author = "Schenke, Bjoern and Venugopalan, Raju",
    title = "{Eccentric protons? Sensitivity of flow to system size and shape in p+p, p+Pb and Pb+Pb collisions}",
    eprint = "1405.3605",
    archivePrefix = "arXiv",
    primaryClass = "nucl-th",
    doi = "10.1103/PhysRevLett.113.102301",
    journal = "Phys. Rev. Lett.",
    volume = "113",
    pages = "102301",
    year = "2014"
}

@article{Yan:2025kwz,
    author = "Yan, Zhengxi",
    title = "{Exploring the Origin of Anisotropy in Small Systems: From Symmetric (O+O) to Asymmetric (d+Au) Collisions}",
    eprint = "2510.03454",
    archivePrefix = "arXiv",
    primaryClass = "nucl-ex",
    month = "10",
    year = "2025"
}

@misc{OSPool,
  doi = {10.21231/906P-4D78},
  url = {https://osg-htc.org/services/open_science_pool.html},
  author = {{OSG}},
  title = {OSPool},
  publisher = {OSG},
  year = {2006}
}

@misc{OSDF,
  doi = {10.21231/0KVZ-VE57},
  url = {https://osdf.osg-htc.org/},
  author = {{OSG}},
  title = {Open Science Data Federation},
  publisher = {OSG},
  year = {2015}
}

@article{Broniowski:2010jd,
    author = "Broniowski, Wojciech and Rybczynski, Maciej",
    title = "{Two-body nucleon-nucleon correlations in Glauber models of relativistic heavy-ion collisions}",
    eprint = "1003.1088",
    archivePrefix = "arXiv",
    primaryClass = "nucl-th",
    doi = "10.1103/PhysRevC.81.064909",
    journal = "Phys. Rev. C",
    volume = "81",
    pages = "064909",
    year = "2010"
}

@article{Lappi:2012vw,
    author = {Lappi, T. and M\"antysaari, H.},
    title = "{On the running coupling in the JIMWLK equation}",
    eprint = "1212.4825",
    archivePrefix = "arXiv",
    primaryClass = "hep-ph",
    doi = "10.1140/epjc/s10052-013-2307-z",
    journal = "Eur. Phys. J. C",
    volume = "73",
    number = "2",
    pages = "2307",
    year = "2013"
}

@article{Cali:2021tsh,
    author = "Cali, Salvatore and Cichy, Krzysztof and Korcyl, Piotr and Kotko, Piotr and Kutak, Krzysztof and Marquet, Cyrille",
    title = "{On systematic effects in the numerical solutions of the JIMWLK equation}",
    eprint = "2104.14254",
    archivePrefix = "arXiv",
    primaryClass = "hep-ph",
    reportNumber = "IFJPAN-IV-2021-7",
    doi = "10.1140/epjc/s10052-021-09380-6",
    journal = "Eur. Phys. J. C",
    volume = "81",
    number = "7",
    pages = "663",
    year = "2021"
}

@article{Giacalone:2025vxa,
    author = "Giacalone, Giuliano and others",
    title = "{Nuclear Physics Confronts Relativistic Collisions Of Isobars}",
    eprint = "2507.01454",
    archivePrefix = "arXiv",
    primaryClass = "nucl-ex",
    month = "7",
    year = "2025"
}

@article{Albacete:2010sy,
    author = "Albacete, Javier L. and Armesto, Nestor and Milhano, Jose Guilherme and Quiroga-Arias, Paloma and Salgado, Carlos A.",
    title = "{AAMQS: A non-linear QCD analysis of new HERA data at small-x including heavy quarks}",
    eprint = "1012.4408",
    archivePrefix = "arXiv",
    primaryClass = "hep-ph",
    doi = "10.1140/epjc/s10052-011-1705-3",
    journal = "Eur. Phys. J. C",
    volume = "71",
    pages = "1705",
    year = "2011"
}

@article{Mueller:2001uk,
    author = "Mueller, Alfred H.",
    title = "{A Simple derivation of the JIMWLK equation}",
    eprint = "hep-ph/0110169",
    archivePrefix = "arXiv",
    reportNumber = "CU-TP-1031",
    doi = "10.1016/S0370-2693(01)01343-0",
    journal = "Phys. Lett. B",
    volume = "523",
    pages = "243--248",
    year = "2001"
}

@article{Mantysaari:2025idf,
    author = {M{\"a}ntysaari, Heikki and Le, Anh Dung},
    title = "{Exclusive $\mathrm{J}/\psi$ production off a dilute proton within a refined hotspot description}",
    eprint = "2509.07480",
    archivePrefix = "arXiv",
    primaryClass = "hep-ph",
    month = "9",
    year = "2025"
}

@misc{cms:OOnote,
    collaboration = "CMS",
    title = "{Measurement of the charged particle nuclear modification factor in oxygen-oxygen collisions with CMS}",
    note = "CMS-PAS-HIN-25-008",
}

@article{Gale:2012rq,
      author         = "Gale, Charles and Jeon, Sangyong and Schenke, Bjorn and
                        Tribedy, Prithwish and Venugopalan, Raju",
      title          = "{Event-by-event anisotropic flow in heavy-ion collisions
                        from combined Yang-Mills and viscous fluid dynamics}",
      journal        = "Phys. Rev. Lett.",
      volume         = "110",
      pages          = "012302",
      doi            = "10.1103/PhysRevLett.110.012302",
      year           = "2013",
      eprint         = "1209.6330",
      archivePrefix  = "arXiv",
      primaryClass   = "nucl-th",
      SLACcitation   = "%%CITATION = ARXIV:1209.6330;%%",
}

@article{Mantysaari:2018zdd,
    author = {M\"antysaari, Heikki and Schenke, Bj\"orn},
    title = "{Confronting impact parameter dependent JIMWLK evolution with HERA data}",
    eprint = "1806.06783",
    archivePrefix = "arXiv",
    primaryClass = "hep-ph",
    doi = "10.1103/PhysRevD.98.034013",
    journal = "Phys. Rev. D",
    volume = "98",
    number = "3",
    pages = "034013",
    year = "2018"
}

@article{McDonald:2023qwc,
    author = "McDonald, Scott and Jeon, Sangyong and Gale, Charles",
    title = "{3+1D initialization and evolution of the glasma}",
    eprint = "2306.04896",
    archivePrefix = "arXiv",
    primaryClass = "hep-ph",
    doi = "10.1103/PhysRevC.108.064910",
    journal = "Phys. Rev. C",
    volume = "108",
    number = "6",
    pages = "064910",
    year = "2023"
}

@article{Muller:2012zq,
    author = "Muller, Berndt and Schukraft, Jurgen and Wyslouch, Boleslaw",
    title = "{First Results from Pb+Pb collisions at the LHC}",
    eprint = "1202.3233",
    archivePrefix = "arXiv",
    primaryClass = "hep-ex",
    reportNumber = "CERN-OPEN-2012-005",
    doi = "10.1146/annurev-nucl-102711-094910",
    journal = "Ann. Rev. Nucl. Part. Sci.",
    volume = "62",
    pages = "361--386",
    year = "2012"
}

@article{Harris:2024aov,
    author = {Harris, John W. and M\"uller, Berndt},
    title = "{\textquotedblleft{}QGP Signatures\textquotedblright{} revisited}",
    doi = "10.1140/epjc/s10052-024-12533-y",
    journal = "Eur. Phys. J. C",
    volume = "84",
    number = "3",
    pages = "247",
    year = "2024"
}

@article{Heinz:2013th,
    author = "Heinz, Ulrich and Snellings, Raimond",
    title = "{Collective flow and viscosity in relativistic heavy-ion collisions}",
    eprint = "1301.2826",
    archivePrefix = "arXiv",
    primaryClass = "nucl-th",
    doi = "10.1146/annurev-nucl-102212-170540",
    journal = "Ann. Rev. Nucl. Part. Sci.",
    volume = "63",
    pages = "123--151",
    year = "2013"
}

@article{Gale:2013da,
    author = "Gale, Charles and Jeon, Sangyong and Schenke, Bjoern",
    title = "{Hydrodynamic Modeling of Heavy-Ion Collisions}",
    eprint = "1301.5893",
    archivePrefix = "arXiv",
    primaryClass = "nucl-th",
    doi = "10.1142/S0217751X13400113",
    journal = "Int. J. Mod. Phys. A",
    volume = "28",
    pages = "1340011",
    year = "2013"
}

@article{Jeon:2015dfa,
    author = "Jeon, Sangyong and Heinz, Ulrich",
    title = "{Introduction to Hydrodynamics}",
    eprint = "1503.03931",
    archivePrefix = "arXiv",
    primaryClass = "hep-ph",
    doi = "10.1142/S0218301315300106",
    journal = "Int. J. Mod. Phys. E",
    volume = "24",
    number = "10",
    pages = "1530010",
    year = "2015"
}

@book{Romatschke:2017ejr,
    author = "Romatschke, Paul and Romatschke, Ulrike",
    title = "{Relativistic Fluid Dynamics In and Out of Equilibrium}",
    eprint = "1712.05815",
    archivePrefix = "arXiv",
    primaryClass = "nucl-th",
    doi = "10.1017/9781108651998",
    isbn = "978-1-108-48368-1, 978-1-108-75002-8",
    publisher = "Cambridge University Press",
    series = "Cambridge Monographs on Mathematical Physics",
    month = "5",
    year = "2019"
}

@article{Miller:2007ri,
    author = "Miller, Michael L. and Reygers, Klaus and Sanders, Stephen J. and Steinberg, Peter",
    title = "{Glauber modeling in high energy nuclear collisions}",
    eprint = "nucl-ex/0701025",
    archivePrefix = "arXiv",
    doi = "10.1146/annurev.nucl.57.090506.123020",
    journal = "Ann. Rev. Nucl. Part. Sci.",
    volume = "57",
    pages = "205--243",
    year = "2007"
}

@article{Schenke:2012wb,
    author = "Schenke, Bjoern and Tribedy, Prithwish and Venugopalan, Raju",
    title = "{Fluctuating Glasma initial conditions and flow in heavy ion collisions}",
    eprint = "1202.6646",
    archivePrefix = "arXiv",
    primaryClass = "nucl-th",
    doi = "10.1103/PhysRevLett.108.252301",
    journal = "Phys. Rev. Lett.",
    volume = "108",
    pages = "252301",
    year = "2012"
}

@article{Schenke:2020mbo,
    author = "Schenke, Bjoern and Shen, Chun and Tribedy, Prithwish",
    title = "{Running the gamut of high energy nuclear collisions}",
    eprint = "2005.14682",
    archivePrefix = "arXiv",
    primaryClass = "nucl-th",
    doi = "10.1103/PhysRevC.102.044905",
    journal = "Phys. Rev. C",
    volume = "102",
    number = "4",
    pages = "044905",
    year = "2020"
}

@article{Iancu:2000hn,
    author = "Iancu, Edmond and Leonidov, Andrei and McLerran, Larry D.",
    title = "{Nonlinear gluon evolution in the color glass condensate. 1.}",
    eprint = "hep-ph/0011241",
    archivePrefix = "arXiv",
    reportNumber = "SACLAY-T00-166, BNL-NT-00-24",
    doi = "10.1016/S0375-9474(01)00642-X",
    journal = "Nucl. Phys. A",
    volume = "692",
    pages = "583--645",
    year = "2001"
}

@article{Schenke:2016ksl,
    author = "Schenke, Bjoern and Schlichting, Soeren",
    title = "{3D glasma initial state for relativistic heavy ion collisions}",
    eprint = "1605.07158",
    archivePrefix = "arXiv",
    primaryClass = "hep-ph",
    doi = "10.1103/PhysRevC.94.044907",
    journal = "Phys. Rev. C",
    volume = "94",
    number = "4",
    pages = "044907",
    year = "2016"
}

@article{Schenke:2022mjv,
    author = "Schenke, Bjoern and Schlichting, Soeren and Singh, Pragya",
    title = "{Rapidity dependence of initial state geometry and momentum correlations in p+Pb collisions}",
    eprint = "2201.08864",
    archivePrefix = "arXiv",
    primaryClass = "nucl-th",
    doi = "10.1103/PhysRevD.105.094023",
    journal = "Phys. Rev. D",
    volume = "105",
    number = "9",
    pages = "094023",
    year = "2022"
}

@inbook{Iancu:2003xm,
    author = "Iancu, Edmond and Venugopalan, Raju",
    editor = "Hwa, Rudolph C. and Wang, Xin-Nian",
    title = "{The Color glass condensate and high-energy scattering in QCD}",
    booktitle = "{Quark-gluon plasma 4}",
    eprint = "hep-ph/0303204",
    archivePrefix = "arXiv",
    doi = "10.1142/9789812795533_0005",
    pages = "249--3363",
    month = "3",
    year = "2003",
note="In \emph{Quark-gluon plasma 4} (eds. R.C. Hwa and X.-N. Wang)"
}

@article{Krasnitz:1998ns,
    author = "Krasnitz, Alex and Venugopalan, Raju",
    title = "{Nonperturbative computation of gluon minijet production in nuclear collisions at very high-energies}",
    eprint = "hep-ph/9809433",
    archivePrefix = "arXiv",
    reportNumber = "NBI-98-21, UALG-TP-98-6",
    doi = "10.1016/S0550-3213(99)00366-1",
    journal = "Nucl. Phys. B",
    volume = "557",
    pages = "237",
    year = "1999"
}

@article{Krasnitz:1999wc,
    author = "Krasnitz, Alex and Venugopalan, Raju",
    title = "{The Initial energy density of gluons produced in very high-energy nuclear collisions}",
    eprint = "hep-ph/9909203",
    archivePrefix = "arXiv",
    doi = "10.1103/PhysRevLett.84.4309",
    journal = "Phys. Rev. Lett.",
    volume = "84",
    pages = "4309--4312",
    year = "2000"
}

@article{Krasnitz:2000gz,
    author = "Krasnitz, Alex and Venugopalan, Raju",
    title = "{The Initial gluon multiplicity in heavy ion collisions}",
    eprint = "hep-ph/0007108",
    archivePrefix = "arXiv",
    doi = "10.1103/PhysRevLett.86.1717",
    journal = "Phys. Rev. Lett.",
    volume = "86",
    pages = "1717--1720",
    year = "2001"
}

@article{Schenke:2010nt,
    author = "Schenke, Bjoern and Jeon, Sangyong and Gale, Charles",
    title = "{(3+1)D hydrodynamic simulation of relativistic heavy-ion collisions}",
    eprint = "1004.1408",
    archivePrefix = "arXiv",
    primaryClass = "hep-ph",
    doi = "10.1103/PhysRevC.82.014903",
    journal = "Phys. Rev. C",
    volume = "82",
    pages = "014903",
    year = "2010"
}

@article{Lappi:2016fmu,
    author = {Lappi, T. and M{\"a}ntysaari, H.},
    title = "{Next-to-leading order Balitsky-Kovchegov equation with resummation}",
    eprint = "1601.06598",
    archivePrefix = "arXiv",
    primaryClass = "hep-ph",
    doi = "10.1103/PhysRevD.93.094004",
    journal = "Phys. Rev. D",
    volume = "93",
    number = "9",
    pages = "094004",
    year = "2016"
}

@article{Paquet:2015lta,
    author = {Paquet, Jean-Fran\c{c}ois and Shen, Chun and Denicol, Gabriel S. and Luzum, Matthew and Schenke, Bj\"orn and Jeon, Sangyong and Gale, Charles},
    title = "{Production of photons in relativistic heavy-ion collisions}",
    eprint = "1509.06738",
    archivePrefix = "arXiv",
    primaryClass = "hep-ph",
    doi = "10.1103/PhysRevC.93.044906",
    journal = "Phys. Rev. C",
    volume = "93",
    number = "4",
    pages = "044906",
    year = "2016"
}

@article{Bazavov:2014pvz,
    author = "Bazavov, A. and others",
    collaboration = "HotQCD",
    title = "{Equation of state in (2+1)-flavor QCD}",
    eprint = "1407.6387",
    archivePrefix = "arXiv",
    primaryClass = "hep-lat",
    reportNumber = "BNL-105928-2014-JA",
    doi = "10.1103/PhysRevD.90.094503",
    journal = "Phys. Rev. D",
    volume = "90",
    pages = "094503",
    year = "2014"
}

@article{Moreland:2015dvc,
    author = "Moreland, J. Scott and Soltz, Ron A.",
    title = "{Hydrodynamic simulations of relativistic heavy-ion collisions with different lattice quantum chromodynamics calculations of the equation of state}",
    eprint = "1512.02189",
    archivePrefix = "arXiv",
    primaryClass = "nucl-th",
    doi = "10.1103/PhysRevC.93.044913",
    journal = "Phys. Rev. C",
    volume = "93",
    number = "4",
    pages = "044913",
    year = "2016"
}

@article{Denicol:2012cn,
    author = "Denicol, G. S. and Niemi, H. and Molnar, E. and Rischke, D. H.",
    title = "{Derivation of transient relativistic fluid dynamics from the Boltzmann equation}",
    eprint = "1202.4551",
    archivePrefix = "arXiv",
    primaryClass = "nucl-th",
    doi = "10.1103/PhysRevD.85.114047",
    journal = "Phys. Rev. D",
    volume = "85",
    pages = "114047",
    year = "2012",
    note = "[Erratum: Phys.Rev.D 91, 039902 (2015)]"
}

@article{Huovinen:2012is,
    author = "Huovinen, Pasi and Petersen, Hannah",
    title = "{Particlization in hybrid models}",
    eprint = "1206.3371",
    archivePrefix = "arXiv",
    primaryClass = "nucl-th",
    doi = "10.1140/epja/i2012-12171-9",
    journal = "Eur. Phys. J. A",
    volume = "48",
    pages = "171",
    year = "2012"
}

@article{Shen:2014vra,
    author = "Shen, Chun and Qiu, Zhi and Song, Huichao and Bernhard, Jonah and Bass, Steffen and Heinz, Ulrich",
    title = "{The iEBE-VISHNU code package for relativistic heavy-ion collisions}",
    eprint = "1409.8164",
    archivePrefix = "arXiv",
    primaryClass = "nucl-th",
    doi = "10.1016/j.cpc.2015.08.039",
    journal = "Comput. Phys. Commun.",
    volume = "199",
    pages = "61--85",
    year = "2016"
}

@misc{ipglasma_code,
  
  title        = {{IP-Glasma}},
  year         = {2024},
  howpublished = {\url{https://github.com/schenke/ipglasma}},
}

@misc{data_link,
  title        = {{Model data link}},
  year         = {2025},
  howpublished = {\url{https://zenodo.org/records/17583744}},
}

@misc{jimwlk_code,
  
  title        = {{JIMWLK}},
  year         = {2022},
  howpublished = {\url{https://github.com/hejajama/jimwlk}},
}

@misc{ipglasma_jimwlk_code,
  
  title        = {{JIMWLK + IP-Glasma}},
  year         = {2025},
  howpublished = {\url{https://github.com/schenke/ipglasma/tree/ipglasma_jimwlk}},
}

@article{Moutarde:2018kwr,
    author = "Moutarde, H. and Sznajder, P. and Wagner, J.",
    title = "{Border and skewness functions from a leading order fit to DVCS data}",
    eprint = "1807.07620",
    archivePrefix = "arXiv",
    primaryClass = "hep-ph",
    doi = "10.1140/epjc/s10052-018-6359-y",
    journal = "Eur. Phys. J. C",
    volume = "78",
    number = "11",
    pages = "890",
    year = "2018"
}

@article{STAR:2008med,
    author = "Abelev, B. I. and others",
    collaboration = "STAR",
    title = "{Systematic Measurements of Identified Particle Spectra in $p p, d^+$ Au and Au+Au Collisions from STAR}",
    eprint = "0808.2041",
    archivePrefix = "arXiv",
    primaryClass = "nucl-ex",
    doi = "10.1103/PhysRevC.79.034909",
    journal = "Phys. Rev. C",
    volume = "79",
    pages = "034909",
    year = "2009"
}

@article{ALICE:2025cjn,
    author = "Acharya, Shreyasi and others",
    collaboration = "ALICE",
    title = "{Centrality dependence of charged-particle pseudorapidity density at midrapidity in Pb-Pb collisions at $\mathbf{\sqrt{\textit{s}_{\rm NN}} = 5.36}$ TeV}",
    eprint = "2504.02505",
    archivePrefix = "arXiv",
    primaryClass = "nucl-ex",
    reportNumber = "CERN-EP-2025-073",
    month = "4",
    year = "2025"
}

@article{ALICE:2016ccg,
    author = "Adam, Jaroslav and others",
    collaboration = "ALICE",
    title = "{Anisotropic flow of charged particles in Pb-Pb collisions at $\sqrt{s_{\rm NN}}=5.02$ TeV}",
    eprint = "1602.01119",
    archivePrefix = "arXiv",
    primaryClass = "nucl-ex",
    reportNumber = "CERN-EP-2016-018",
    doi = "10.1103/PhysRevLett.116.132302",
    journal = "Phys. Rev. Lett.",
    volume = "116",
    number = "13",
    pages = "132302",
    year = "2016"
}

@article{STAR:2017idk,
    author = "Adamczyk, L. and others",
    collaboration = "STAR",
    title = "{Harmonic decomposition of three-particle azimuthal correlations at energies available at the BNL Relativistic Heavy Ion Collider}",
    eprint = "1701.06496",
    archivePrefix = "arXiv",
    primaryClass = "nucl-ex",
    doi = "10.1103/PhysRevC.98.034918",
    journal = "Phys. Rev. C",
    volume = "98",
    number = "3",
    pages = "034918",
    year = "2018"
}

@article{STAR:2016vqt,
    author = "Adamczyk, L. and others",
    collaboration = "STAR",
    title = "{Beam Energy Dependence of the Third Harmonic of Azimuthal Correlations in Au+Au Collisions at RHIC}",
    eprint = "1601.01999",
    archivePrefix = "arXiv",
    primaryClass = "nucl-ex",
    doi = "10.1103/PhysRevLett.116.112302",
    journal = "Phys. Rev. Lett.",
    volume = "116",
    number = "11",
    pages = "112302",
    year = "2016"
}

@article{ALICE:2019zfl,
    author = "Acharya, Shreyasi and others",
    collaboration = "ALICE",
    title = "{Investigations of Anisotropic Flow Using Multiparticle Azimuthal Correlations in pp, p-Pb, Xe-Xe, and Pb-Pb Collisions at the LHC}",
    eprint = "1903.01790",
    archivePrefix = "arXiv",
    primaryClass = "nucl-ex",
    reportNumber = "CERN-EP-2019-033",
    doi = "10.1103/PhysRevLett.123.142301",
    journal = "Phys. Rev. Lett.",
    volume = "123",
    number = "14",
    pages = "142301",
    year = "2019"
}

@article{ATLAS:2019pvn,
    author = "Aad, Georges and others",
    collaboration = "ATLAS",
    title = "{Measurement of flow harmonics correlations with mean transverse momentum in lead-lead and proton-lead collisions at $\sqrt{s_{NN}}=5.02$ TeV with the ATLAS detector}",
    eprint = "1907.05176",
    archivePrefix = "arXiv",
    primaryClass = "nucl-ex",
    reportNumber = "CERN-EP-2019-130",
    doi = "10.1140/epjc/s10052-019-7489-6",
    journal = "Eur. Phys. J. C",
    volume = "79",
    number = "12",
    pages = "985",
    year = "2019"
}

@article{STAR:2019zaf,
    author = "Adam, Jaroslav and others",
    collaboration = "STAR",
    title = "{Azimuthal Harmonics in Small and Large Collision Systems at RHIC Top Energies}",
    eprint = "1901.08155",
    archivePrefix = "arXiv",
    primaryClass = "nucl-ex",
    doi = "10.1103/PhysRevLett.122.172301",
    journal = "Phys. Rev. Lett.",
    volume = "122",
    number = "17",
    pages = "172301",
    year = "2019"
}

@article{lhcb_smog,
    collaboration = "LHCb",
    title = "{LHCb SMOG Upgrade}",
    journal = "{CERN-LHCC-2019-005, LHCB-TDR-020}",
    doi = "10.17181/CERN.SAQC.EOWH"
}

@article{Mantysaari:2025ltq,
    author = {M{\"a}ntysaari, Heikki and Roch, Hendrik and Salazar, Farid and Schenke, Bj{\"o}rn and Shen, Chun and Zhao, Wenbin},
    title = "{Global Bayesian Analysis of $\mathrm{J}/\psi$ Photoproduction on Proton and Lead Targets}",
    eprint = "2507.14087",
    archivePrefix = "arXiv",
    primaryClass = "hep-ph",
    month = "7",
    year = "2025"
}

@article{Mantysaari:2025tcg,
    author = {M{\"a}ntysaari, Heikki and Schenke, Bj{\"o}rn and Shen, Chun and Zhao, Wenbin},
    title = "{Collision-Energy Dependence in Heavy-Ion Collisions from Nonlinear QCD Evolution}",
    eprint = "2502.05138",
    archivePrefix = "arXiv",
    primaryClass = "nucl-th",
    doi = "10.1103/gf4y-p5j7",
    journal = "Phys. Rev. Lett.",
    volume = "135",
    number = "2",
    pages = "022302",
    year = "2025"
}

@article{Rezaeian:2012ji,
    author = "Rezaeian, Amir H. and Siddikov, Marat and Van de Klundert, Merijn and Venugopalan, Raju",
    title = "{Analysis of combined HERA data in the Impact-Parameter dependent Saturation model}",
    eprint = "1212.2974",
    archivePrefix = "arXiv",
    primaryClass = "hep-ph",
    doi = "10.1103/PhysRevD.87.034002",
    journal = "Phys. Rev. D",
    volume = "87",
    number = "3",
    pages = "034002",
    year = "2013"
}

@article{Lappi:2013zma,
    author = {Lappi, T. and M{\"a}ntysaari, H.},
    title = "{Single inclusive particle production at high energy from HERA data to proton-nucleus collisions}",
    eprint = "1309.6963",
    archivePrefix = "arXiv",
    primaryClass = "hep-ph",
    doi = "10.1103/PhysRevD.88.114020",
    journal = "Phys. Rev. D",
    volume = "88",
    pages = "114020",
    year = "2013"
}

@article{Kurkela:2018wud,
    author = {Kurkela, Aleksi and Mazeliauskas, Aleksas and Paquet, Jean-Fran{\c{c}}ois and Schlichting, S{\"o}ren and Teaney, Derek},
    title = "{Matching the Nonequilibrium Initial Stage of Heavy Ion Collisions to Hydrodynamics with QCD Kinetic Theory}",
    eprint = "1805.01604",
    archivePrefix = "arXiv",
    primaryClass = "hep-ph",
    doi = "10.1103/PhysRevLett.122.122302",
    journal = "Phys. Rev. Lett.",
    volume = "122",
    number = "12",
    pages = "122302",
    year = "2019"
}

@article{STAR:2025ivi,
    collaboration = "STAR",
    title = "{Engineering the shapes of quark-gluon plasma droplets by comparing anisotropic flow in small symmetric and asymmetric collision systems}",
    eprint = "2510.19645",
    archivePrefix = "arXiv",
    primaryClass = "nucl-ex",
    month = "10",
    year = "2025"
}

@article{STAR:2015mki,
    author = "Adamczyk, L. and others",
    collaboration = "STAR",
    title = "{Azimuthal anisotropy in U$+$U and Au$+$Au collisions at RHIC}",
    eprint = "1505.07812",
    archivePrefix = "arXiv",
    primaryClass = "nucl-ex",
    doi = "10.1103/PhysRevLett.115.222301",
    journal = "Phys. Rev. Lett.",
    volume = "115",
    number = "22",
    pages = "222301",
    year = "2015"
}

@misc{ALICE:MartaUrioni_IS2025,
    author = { Marta Urioni
},
title={{\emph{Charged particle production at mid and forward rapidities from small to large collision systems with ALICE}}},
    note="Presented at The 8th International Conference on the Initial Stages in High-Energy Nuclear Collisions (Initial Stages 2025)",
url="https://indico.cern.ch/event/1479384/contributions/6663047/attachments/3131557/5555439"
}

@article{Zhao:2022ugy,
    author = {Zhao, Wenbin and Ryu, Sangwook and Shen, Chun and Schenke, Bj\"orn},
    title = "{3D structure of anisotropic flow in small collision systems at energies available at the BNL Relativistic Heavy Ion Collider}",
    eprint = "2211.16376",
    archivePrefix = "arXiv",
    primaryClass = "nucl-th",
    doi = "10.1103/PhysRevC.107.014904",
    journal = "Phys. Rev. C",
    volume = "107",
    number = "1",
    pages = "014904",
    year = "2023"
}

@article{Bass:1998ca,
    author = "Bass, S. A. and others",
    title = "{Microscopic models for ultrarelativistic heavy ion collisions}",
    eprint = "nucl-th/9803035",
    archivePrefix = "arXiv",
    doi = "10.1016/S0146-6410(98)00058-1",
    journal = "Prog. Part. Nucl. Phys.",
    volume = "41",
    pages = "255--369",
    year = "1998"
}

@article{Bleicher:1999xi,
    author = "Bleicher, M. and others",
    title = "{Relativistic hadron hadron collisions in the ultrarelativistic quantum molecular dynamics model}",
    eprint = "hep-ph/9909407",
    archivePrefix = "arXiv",
    doi = "10.1088/0954-3899/25/9/308",
    journal = "J. Phys. G",
    volume = "25",
    pages = "1859--1896",
    year = "1999"
}

@article{Pordes:2007zzb,
    author = "Pordes, Ruth and others",
    editor = "Keyes, David E.",
    title = "{The Open Science Grid}",
    reportNumber = "FERMILAB-CONF-07-217-CD",
    doi = "10.1088/1742-6596/78/1/012057",
    journal = "J. Phys. Conf. Ser.",
    volume = "78",
    pages = "012057",
    year = "2007"
}

@article{Sfiligoi:2009cct,
    author = "Sfiligoi, Igor and Bradley, Daniel C. and Holzman, Burt and Mhashilkar, Parag and Padhi, Sanjay and Wurthwrin, Frank",
    title = "{The pilot way to Grid resources using glideinWMS}",
    reportNumber = "FERMILAB-CONF-09-373-CD",
    doi = "10.1109/CSIE.2009.950",
    journal = "WRI World Congress",
    volume = "2",
    pages = "428--432",
    year = "2009"
}

@article{Mantysaari:2016ykx,
    author = {M\"antysaari, Heikki and Schenke, Bj\"orn},
    title = "{Evidence of strong proton shape fluctuations from incoherent diffraction}",
    eprint = "1603.04349",
    archivePrefix = "arXiv",
    primaryClass = "hep-ph",
    doi = "10.1103/PhysRevLett.117.052301",
    journal = "Phys. Rev. Lett.",
    volume = "117",
    number = "5",
    pages = "052301",
    year = "2016"
}

@article{Carlson:1997qn,
    author = "Carlson, J. and Schiavilla, R.",
    title = "{Structure and Dynamics of Few Nucleon Systems}",
    reportNumber = "JLAB-THY-97-20",
    doi = "10.1103/RevModPhys.70.743",
    journal = "Rev. Mod. Phys.",
    volume = "70",
    pages = "743--842",
    year = "1998"
}

@article{PHENIX:2004vdg,
    author = "Adler, S. S. and others",
    collaboration = "PHENIX",
    title = "{Systematic studies of the centrality and $\sqrt{s_{NN}}$ dependence of the $\mathrm{d}E_T/\mathrm{d}\eta$ and $\mathrm{d}N_\mathrm{ch}/\mathrm{d}\eta$ in heavy ion collisions at mid-rapidity}",
    eprint = "nucl-ex/0409015",
    archivePrefix = "arXiv",
    doi = "10.1103/PhysRevC.71.034908",
    journal = "Phys. Rev. C",
    volume = "71",
    pages = "034908",
    year = "2005",
    note = "[Erratum: Phys.Rev.C 71, 049901 (2005)]"
}

@article{H1:2009pze,
    author = "Aaron, F. D. and others",
    collaboration = "H1, ZEUS",
    title = "{Combined Measurement and QCD Analysis of the Inclusive $e^\pm p$ Scattering Cross Sections at HERA}",
    eprint = "0911.0884",
    archivePrefix = "arXiv",
    primaryClass = "hep-ex",
    reportNumber = "DESY-09-158",
    doi = "10.1007/JHEP01(2010)109",
    journal = "JHEP",
    volume = "01",
    pages = "109",
    year = "2010"
}

@article{Mantysaari:2024qmt,
    author = {M\"antysaari, Heikki and Singh, Pragya},
    title = "{Energy dependence of the deformed nuclear structure at small-$x$}",
    eprint = "2411.14934",
    archivePrefix = "arXiv",
    primaryClass = "nucl-th",
    month = "11",
    year = "2024"
}

@article{Schenke:2010rr,
    author = "Schenke, Bjorn and Jeon, Sangyong and Gale, Charles",
    title = "{Elliptic and triangular flow in event-by-event (3+1)D viscous hydrodynamics}",
    eprint = "1009.3244",
    archivePrefix = "arXiv",
    primaryClass = "hep-ph",
    doi = "10.1103/PhysRevLett.106.042301",
    journal = "Phys. Rev. Lett.",
    volume = "106",
    pages = "042301",
    year = "2011"
}

@inproceedings{Brewer:2021kiv,
    author = "Brewer, Jasmine and Mazeliauskas, Aleksas and van der Schee, Wilke",
    title = "{Opportunities of OO and $p$O collisions at the LHC}",
    booktitle = "{Opportunities of OO and pO collisions at the LHC}",
    eprint = "2103.01939",
    archivePrefix = "arXiv",
    primaryClass = "hep-ph",
    reportNumber = "CERN-TH-2021-028",
    month = "3",
    year = "2021"
}

@article{McLerran:1993ni,
    author = "McLerran, Larry D. and Venugopalan, Raju",
    title = "{Computing quark and gluon distribution functions for very large nuclei}",
    eprint = "hep-ph/9309289",
    archivePrefix = "arXiv",
    reportNumber = "TPI-MINN-93-44-T, NUC-MINN-93-24-T, HEP-UMN-TH-1220-93",
    doi = "10.1103/PhysRevD.49.2233",
    journal = "Phys. Rev. D",
    volume = "49",
    pages = "2233--2241",
    year = "1994"
}

@article{Schlichting:2014ipa,
    author = {Schlichting, S\"oren and Schenke, Bj\"orn},
    title = "{The shape of the proton at high energies}",
    eprint = "1407.8458",
    archivePrefix = "arXiv",
    primaryClass = "hep-ph",
    doi = "10.1016/j.physletb.2014.10.068",
    journal = "Phys. Lett. B",
    volume = "739",
    pages = "313--319",
    year = "2014"
}

\end{document}